\documentclass[journal=jacsat,manuscript=article]{achemso}

\usepackage[version=3]{mhchem} % Formula subscripts using \ce{}
\usepackage{amssymb}
\usepackage{amsmath}
\usepackage{graphicx}

\author{Bryce Rives}
\affiliation[UCB]{Department of Chemical and Biological Engineering, University of Colorado Boulder, Boulder, CO 80303, USA}

\author{Filipe Henrique}
\affiliation[PU]{Department of Mechanical and Aerospace Engineering, Princeton University, NJ, USA}

\author{Pawe\l{} J. \.Zuk}
\affiliation[IPCPAS]{Institute of Physical Chemistry, Polish Academy of Sciences, Warsaw, Poland}

\author{Ankur Gupta}
\email{ankur.gupta@colorado.edu}
\affiliation[UCB]{Department of Chemical and Biological Engineering, University of Colorado Boulder, Boulder, CO 80303, USA}

\title[An \textsf{achemso} demo]
  {Effect of Nanopore Wall Geometry on Electrical Double-Layer Charging Dynamics}


\begin{document}

\iffalse
\begin{tocentry}

Some journals require a graphical entry for the Table of Contents.
This should be laid out ``print ready'' so that the sizing of the
text is correct.

Inside the \texttt{tocentry} environment, the font used is Helvetica
8\,pt, as required by \emph{Journal of the American Chemical
Society}.

The surrounding frame is 9\,cm by 3.5\,cm, which is the maximum
permitted for  \emph{Journal of the American Chemical Society}
graphical table of content entries. The box will not resize if the
content is too big: instead it will overflow the edge of the box.

This box and the associated title will always be printed on a
separate page at the end of the document.

\end{tocentry}
\fi

\begin{abstract}
Confinement strongly influences electrochemical systems, where structural control has enabled advances in nanofluidics, sensing, and energy storage. In electric double-layer capacitors (EDLCs), or supercapacitors, energy density is governed by the accessible surface area of porous electrodes. Continuum models, built on first-principles transport equations, have provided critical insight into electrolyte dynamics under confinement but have largely focused on pores with straight walls. In such geometries, a fundamental trade-off emerges: wider pores charge faster but store less energy, while narrower pores store more charge but charge slowly. Here, we apply perturbation analysis to the Poisson–Nernst–Planck (PNP) equations for a single pore of gradually varying radius, focusing on the small potential and slender aspect ratio regime. Our analysis reveals that sloped pore walls induce an additional ionic flux, enabling simultaneous acceleration of charging and enhancement of charge storage. The theoretical predictions closely agree with direct numerical simulations while reducing computational cost by 5–6 orders of magnitude. We further propose a modified effective circuit representation that captures geometric variation along the pore and demonstrate how the framework can be integrated into pore-network models. This work establishes a scalable approach to link pore geometry with double-layer dynamics and offers new design principles for optimizing supercapacitor performance.
\end{abstract}

\section{Introduction}
Confinement is common in electrochemical systems, and careful control of structure in these systems has enabled breakthroughs in nanofluidics, single-molecule sensing, and enhancements in energy storage devices~\cite{wang2016dynamics, itoi2011three, boyd2021effects}. Focusing on energy storage devices, electric double-layer capacitors (EDLCs), or supercapacitors, offer high power density along with minimal cyclical degradation~\cite{simon2008materials, simon2020perspectives}. EDLCs have diverse current and potential applications---including flexible self-powered electronics, balancing consumption/generation within the electric grid, and powering electric vehicles~\cite{li20203d, li20233d,navarro2021present, presser2012electrochemical,horn2019supercapacitors}---but are limited by their energy densities. Advancements in room temperature ionic liquids (RTILs), and their corresponding theory, have yielded improved understanding with increased performance and safety with their extended voltage window, thermal stability, and nonflammable nature~\cite{abdallah2012room,watanabe2017application,azov2018solvent, kondrat2023theory,lee2015dynamics}. Since the performance of EDLCs is proportional to surface area, porous carbon electrodes are employed to enhance energy storage~\cite{simon2020perspectives}. This raises a key question: how do we optimize pore design inside electrodes? To this end, an understanding of electrolyte transport under charged confinement is crucial to obtain structure-property relationships and optimize performance. 
\par{} Molecular dynamics (MD) simulations are commonly used to model electrolyte-electrode interactions under confinement---offering insights into atomic-scale phenomena---since pore length scales are often comparable to the ionic radius. Recently, MD studies have revealed insights into the dynamics of ionic liquids~\cite{mo2023molecular, bi2020molecular}, optimal charging--discharging cycles~\cite{breitsprecher2018charge, breitsprecher2020speed}, the importance of gradual desolvation in nanopores~\cite{mo2023horn}, and benefits of ionophobic pore walls~\cite{kondrat2014accelerating, gan2021ionophobic}. Although MD simulations are able to capture atomic-level resolution inside nanopores, their computational expense typically limits their simulation size to a single nanopore and short timescales. Hence, parallel advances are required in continuum modeling to predict electrolyte transport in complex geometries and at long timescales.
\par{} At the continuum level, de Levie pioneered modeling EDL formation within porous electrodes with an equivalent circuit model~\cite{de1963porous, de1964porous}, which remains influential today. In the past few years, there have been exciting developments in this area, such as the inclusion of nonlinear potentials~\cite{biesheuvel2010nonlinear}, Faradaic reactions~\cite{biesheuvel2011diffuse}, surface conduction~\cite{mirzadeh2014enhanced}, arbitrary Debye lengths~\cite{henrique2022charging}, diffusivity asymmetry~\cite{henrique2022impact, henrique2025parallel}, floating electrodes~\cite{} concentrated electrolytes~\cite{fertig2025charging}, impact of convection~\cite{ratschow2024convection} and a network of pores~\cite{henrique2024network, alizadeh2017multiscale, alizadeh2019impact}. Notably, some of these approaches model a macrohomogeneous electrode, representing the electrode collective behavior rather than at the pore scale. These models can produce equivalent circuits similar to de Levie transmission line models; however, the underlying assumptions and starting points of these approaches are different. While these models provide deep insights into the charging of double-layer inside porous materials, many studies focus on cylindrical or slit-pore~\cite{aslyamov2022analytical, yang2022simulating, pedersen2023equivalent} shapes with a constant radius or height, i.e., straight walls. One key result that these studies reveal is that for small to moderate potentials, wider pores charge faster at the expense of a lower charge density~\cite{henrique2022charging}. Here, we address the question: is it possible to use sloped walls to enhance the charging rates, while also increasing the density of charge stored? Our results reveal, rather surprisingly, that this is indeed possible due to the additional flux that the sloped walls create.
\par{} By building on our prior work~\cite{henrique2022charging, henrique2022impact, henrique2024network, henrique2025parallel}, we employ perturbation analysis to develop a theoretical model to predict double-layer charging for an axisymmetric pore with gradually varying radius, under the long-slender pore assumption. The theoretical model reveals that sloped walls create an additional flux for all Debye lengths, which can be used to speed up (or slow down) the charging of double-layers. The model is in excellent agreement with direct numerical simulations and can successfully capture the spatiotemporal variations of charges and potential with a 5–6 order-of-magnitude reduction in computational cost. We also proposed a modified effective circuit that is able to capture geometrical variations along the pore. Finally, given that our theory is computationally friendly, it can easily be integrated with network models to simulate thousands of interconnected pores~\cite{henrique2024network, alizadeh2017multiscale, alizadeh2019impact} and provides a crucial step towards obtaining structure-property relationships in supercapacitors to optimize performance.  

\section{Results and discussion}
\begin{figure}[t!]
    \centering
    \includegraphics[width=0.8\linewidth]{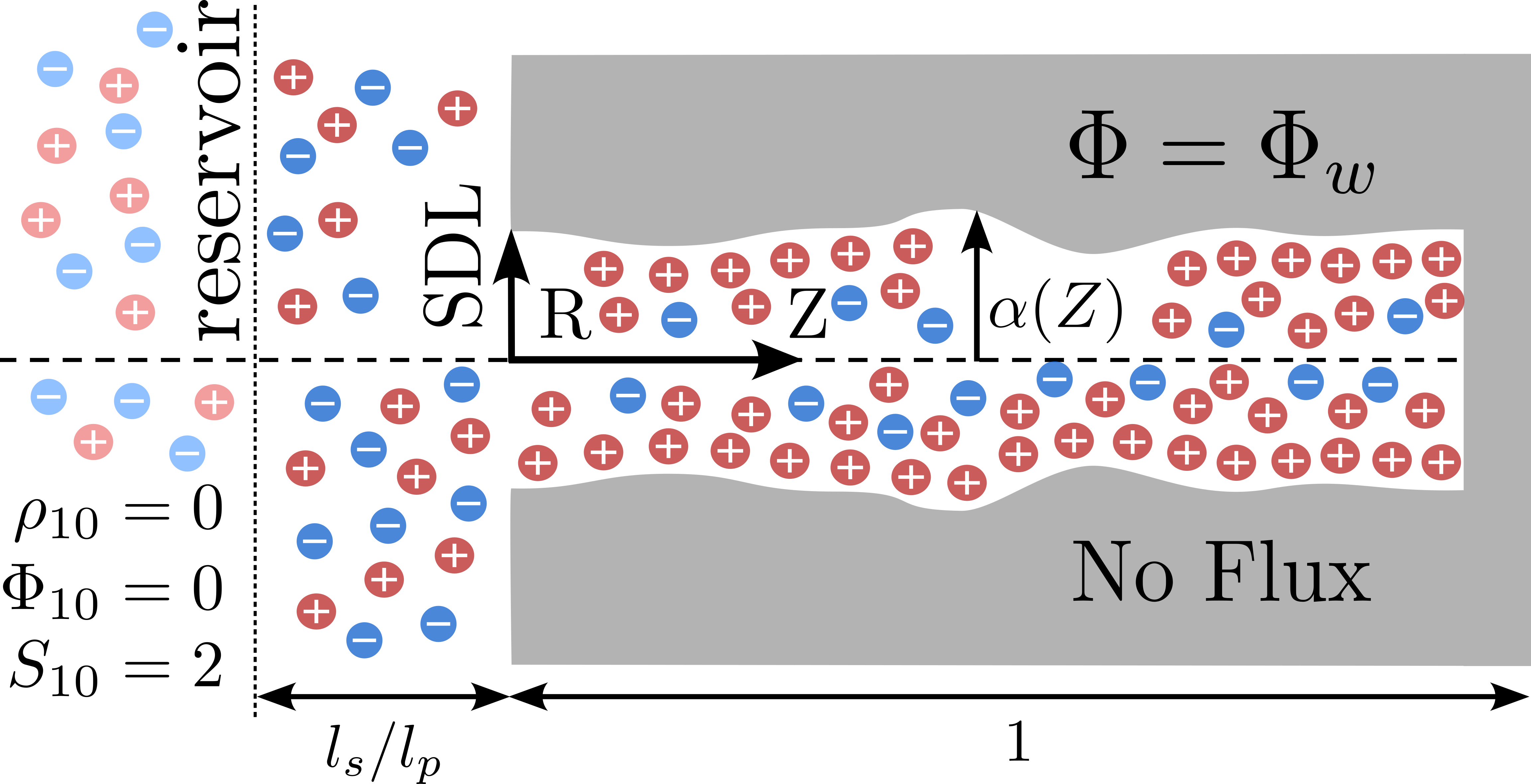}
    \caption{
    A schematic of generalized nanopore, with red circles representing cations and blue circles representing anions. Although initially uncharged, at time $\tau=0$, a negative potential of {$\Phi_w$} is applied to the pore wall, driving cations into the pore and expelling anions. There exist three distinct regions: the reservoir, the static diffusion layer, and the pore. The reservoir acts as an unchanging repository of charged ions, allowing ions to flow in and out freely. The reservoir  remains electroneutral, serving as the reference level for potential. The static diffusion layer (SDL) serves as a bridge between the reservoir and the pore. While electrically neutral, its potential increases linearly as ions approach the pore entrance. The radius and length scale of this region are adjustable parameters, when formulated appropriately, define the dimensionless Biot number---a quantity characterizing the resistance of electrochemical flux from the reservoir region to the pore entrance. The pore region is where the EDL forms, and is the main focus of analysis. Importantly, the walls are ideally blocking, and the pore’s axial length is much larger than its radial dimension. The  radius of the pore depends on the axial coordinate, given by $\alpha(Z)$.}
    \label{Fig: schematic}
\end{figure}

\subsection{Problem setup}
We consider an idealized axisymmetric single pore where the pore radius changes along the axial direction; see Fig.~\ref{Fig: schematic}. We assume the electrolyte to be binary, monovalent with valences $z_{\pm}=\pm 1$ and symmetric diffusivities such that $D_{\pm}=D$. The concentration of the cation is described by $c_+$; conversely, the concentration of the anion is described by $c_-$. We ignore surface redox reactions and assume the surface to be ideally blocking, i.e., ion fluxes through a surface are negligible. 
\par{}  We assume there exist three distinct regions, see Fig.~\ref{Fig: schematic}: the reservoir, the static diffusion layer, and the pore~\cite{biesheuvel2010nonlinear,henrique2022charging,henrique2024network,gupta2020PRL,janssen2021transmission,khair2008fundamental,henrique2022impact}. Far from the electrode surface, the reservoir consists of ions with concentration $c_\pm=c_0$ and potential $\phi=0$. This reference potential---chosen for mathematical convenience---is ultimately arbitrary and different system perspectives exist; for instance, a 1D two-electrodes system would have a point reference potential~\cite{lian2020blessing}. An electroneutral static diffusion layer (SDL) forms between the pore entrance and the reservoir~\cite{biesheuvel2010nonlinear,henrique2022charging,henrique2024network,gupta2020PRL,henrique2022impact}. Although the pore region---where ions accumulate to form EDLs---is the main focus of this analysis, the SDL is meant to capture transport limitations (entrance resistance) outside the pore. %  We highlight several important effects that the entrance resistance has on charging rates across several geometries, noting that entrance effects outside the pore remain an underexplored area of research. 
The SDL and pore region have associated length scales $\ell_s^*$ and $\ell_p^*$ respectively, as well as radii $a_s^*$ and $a_p^*$---noting that $(^*)$ terms are dimensional quantities. We assume the pore length scale to be much greater than the radial length scale, such that $a_p^*\ll\ell_p^*$. We write the shape equation of the pore as
\begin{equation}
    r= \alpha^*(z) ,\label{Eq: dimensionalPoreWall}
\end{equation} 
where $\alpha^*(z)$ is the local radius, which changes along the $z-$direction. We assume that the length scales of the SDL and pore region are of the same order, hence $\ell_s^*/\ell_p^* = O(1)$. 
\par{} Initially, the concentration everywhere inside SDL and the pore is assumed to be the same as the concentration of the reservoir, $c_{\pm}=c_0$. To initiate charging, we set the pore surfaces to be a constant potential $\phi = \Phi_w^*$ at $t>0$, where $t$ is time. To illustrate the physical process, we assume the potential to be negative. Consequently, cations will preferentially transport into the pore and anions will transport out of the pore, forming a double-layer. The double-layer will form initially at the pore mouth only. As time progresses, ions will continue to flux in and out of the pore, developing the double-layer further into the pore until equilibrium is reached. Our objective is to quantify this dynamic process and especially elucidate the effect of pore geometry.      
\subsection{Derivation outline}
\par{} Due to the length of the derivation section, we provide a brief summary here and highlight the key assumptions of our work. Additionally, as many different symbols are used, a notation table is provided in the Supplementary Information (Table~\ref{tab:symbols}). Our goal is to capture the transient charging behavior within a single shape-changing pore. We use the Planck–Nernst–Poisson (PNP) equations to model ion fluxes within the pore. We apply a two-term perturbation expansion in the small potential and slender aspect ratio regimes to linearize the PNP equations. From this expansion, we derive the electrochemical potential of charge and find that it is radially constant. This quantity, which we will later define, allows us to recover both the potential and charge density profiles within the pore. After radial averaging, we recover a transient diffusion equation that depends only on the axial position. For the boundary conditions used to solve our PDE, we assume the walls to be ideally blocking and the pore to be connected to an electrically neutral region---the static diffusion layer---where a flux-matching condition links the two domains. This electroneutral region introduces additional parameters that account for transport inefficiency from the reservoir to pore region.
\subsection{Governing equations}
\par{} We first focus on the pore region and capture the SDL afterward. We assume that the ion transport inside the pore is described by the Nernst-Planck equation, which is given as
\begin{subequations}
\label{Eq: geqn_dim}
\begin{equation}
    \frac{\partial c_{\pm}}{\partial t}+\nabla\cdot \mathbf{N}_{\pm}=0,
    \label{Eq: NP}
\end{equation}
where the ion fluxes $\mathbf{N}_\pm$ are given as
\begin{equation}
    \mathbf{N}_{\pm}=-D\nabla c_{\pm}\mp\frac{De}{k_B T}c_{\pm}\nabla\phi,
\end{equation}
where $\phi$ is the electrical potential, $e$ is the fundamental charge, $k_B$ is the Boltzmann constant and $T$ is the absolute temperature. The first term of the RHS is the diffusive flux and the second term is the electromigrative flux. We note that several other modifications of the Poisson-Nernst-Planck exist, including finite ion-size~\cite{kilic2007steric,lee2015dynamics,gupta2018electrical, lee2015dynamics}, dielectric decrement~\cite{nakayama2015differential,gupta2018electrical} and ion-ion correlations~\cite{nakayama2015differential,gupta2018electrical,de2020continuum,zheng2025diffusive,gavish2018solvent}, but are not considered for simplicity. While these effects will modify our results, we anticipate that the qualitative influence of geometry is model-independent; however, the quantitative impacts may vary, and a full analysis of the modified equations will be required to determine its magnitude. We also note that recently, convection has been argued to be important in the pore charging process~\cite{ratschow2024convection}, but it is not included here. This is consistent with the linearized regime we investigate, as explained later. However, interested readers are referred to~\cite{ratschow2024convection} for the effect of convection in the nonlinear regime. 
\par{} To complete the model, we invoke  the Poisson equation,
\begin{equation}
    -\epsilon\nabla^2\phi=e(c_+-c_-),
    \label{Eq: poisson}
\end{equation}
\end{subequations}
where $\epsilon$ is the electrical permittivity of the electrolyte solution, which is treated as constant in this work, but $\epsilon$ can depend on ion concentration due to dielectric decrement effects~\cite{nakayama2015differential, gupta2018electrical}.
\subsection{Non-dimensional equations}
For simplification, we nondimensionalize by the cation and anion reservoir concentration, $c_0$, into a charge density, $\rho=\frac{c_+-c_-}{c_0}$, and a salt concentration, $S=\frac{c_++c_-}{c_0}$. The potential is normalized by the latent thermal voltage, $\Phi=\frac{e\phi}{k_BT}$. The time is non-dimensionalized by the diffusion time $\tau = \frac{D t}{\ell_p^{*2}}$. The Debye length is given as $\lambda=\sqrt{\frac{\epsilon k_BT}{2e^2c_0}}$ and non-dimensional Debye ratio $\kappa=\frac{a_p^*}{\lambda}$, where $a_p^*$ is a reference pore radius. The two different length scales (radial and axial) are used to non-dimensionalize the length, so $R=\frac{r}{a_p^*}$ and $Z=\frac{z}{\ell_p^*}$. We also define $\alpha(Z) = \frac{\alpha^*}{a_p^*}$. Converting Eqs. \eqref{Eq: geqn_dim} in terms of these non-dimensional parameters,
\begin{subequations}
\label{Eq: geqn_nondim}
\begin{eqnarray}
    \frac{\partial\rho}{\partial \tau}=\boldsymbol{\nabla}^2\rho+\boldsymbol{\nabla}\cdot(S\boldsymbol{\nabla}\Phi),\label{Eq: rhoNP} \\
    \frac{\partial S}{\partial \tau} = \boldsymbol{\nabla}^2 S+\boldsymbol{\nabla}\cdot(\rho \boldsymbol{\nabla}\Phi),\label{Eq: saltNP}  \\
    \frac{a_p^{*2}}{\ell_p^{*2}} \boldsymbol{\nabla}^2\Phi=-\frac{\kappa^2}{2} \rho, \label{Eq: poisson}
\end{eqnarray}
\end{subequations}
where $\boldsymbol \nabla =\frac{\ell_p^*}{a_p^*} \frac{\partial}{\partial R} \mathbf{e}_R  + \frac{\partial}{\partial Z} \mathbf{e}_Z$ and $(^*)$ denotes a dimensional length. We introduce the reference radius, $a_p^*$, to enable relative comparison between pores of different sizes. For example, if pore A is twice as large as pore B, then pore A's non-dimensional wall equation would correspond to $2\alpha(Z)$ given the same $a_p^*$. Although other normalization schemes can be used, we keep $a_p^*$ constant between all pores and simulations for simplicity. % use this reference to preserve the proportional sizes of different pores. Therefore, by defining $a_p^*$ differently, keep the relative magnitude of the Debye ratio constant, $\kappa=\frac{a_p^*}{\lambda}$, while preserving the proportional sizes of different pores
\subsection{Perturbation expansion}
To make progress in solving Eqs. \eqref{Eq: geqn_nondim}, we consider a two-term perturbation expansion for the parameters $\Phi_w$ and $\delta$, where both $\Phi_w \ll 1 $ and $\delta \ll 1$. The first parameter, $\Phi_w$ corresponds to a small applied wall potential, and our prior work has shown that a linearized solution compares favorably to direct numerical simulation up to $\Phi_w \approx 4$~\cite{henrique2022charging}. The second parameter, $\delta$, represents the slender aspect ratio, defined as $\delta = a_p^{*2}/l_p^{*2} \ll 1$. Beyond restricting the pore's aspect ratio, $\delta$ imposes further constraints on pore wall variations, which are quantified later on. The regular perturbation expansions are then proposed as follows:
\begin{subequations}
\begin{eqnarray}
\rho=\rho_{00}+\Phi_w (\rho_{10}+\delta \rho_{11}+\ldots)+\Phi_w^2 (\rho_{20}+\delta \rho_{21}+\ldots)+\ldots \\
\Phi=\Phi_{00}+\Phi_w (\Phi_{10}+\delta \Phi_{11}+\ldots)+\Phi_w^2 (\Phi_{20}+\delta \Phi_{21}+\ldots)+\ldots \\
S=S_{00}+\Phi_w (S_{10}+\delta S_{11}+\ldots)+\Phi_w^2 (S_{20}+\delta S_{21}+\ldots)+\ldots 
\end{eqnarray}
\end{subequations}
Here, the first subscript denotes the order of the solution in $\Phi_w$, and the second subscript denotes the order in $\delta$, e.g. $\rho_{21}$ corresponds to $O(\Phi_w^2\delta^1)$. Although not immediately evident, except for $S_{00}$, all $O(\Phi_w^0\delta^n)$ terms vanish. In the absence of an applied electric field, there is no driving force for ions to move in or out of the pore, and thus $O(\Phi_w^0\delta^n)$ corrections are not included in the expansion. The inclusion of $\delta$ is vital, as it enables a lubrication-like approximation that accounts for variations in pore radius.
\par{} The leading-order problem is trivial to solve, as it corresponds to a binary electrolyte with no applied potential. Consequently, no charge density or potential gradient develops within the electrolyte, yielding $\rho_{00}(R,Z,\tau)=\Phi_{00}(R,Z,\tau)=0$. The ion concentration remains constant and equal to that of the reservoir concentration; hence, the leading-order salt concentration $S_{00}(R,Z,\tau)=2$.
\par{} A careful examination of the next order reveals that radial scaling plays a crucial role in determining which terms appear. In particular, due to geometric scaling in front of the gradient operator, Eq.~\eqref{Eq: rhoNP} has a term at $O\left(\Phi_w\delta^{-1}\right)$, which reads
\begin{subequations}
\begin{equation}
    \frac{\partial}{\partial R}\left(R\frac{\partial}{\partial R} (\rho_{10}+2\Phi_{10})\right)=0,
\end{equation}
after integrating and employing symmetry at $R=0$ becomes
\begin{equation}
\frac{\partial}{\partial R} \left(\rho_{10}+2\Phi_{10} \right) = 0.
\label{Eq: symmetry}
\end{equation}
\label{Eq: del_minus_1}
\end{subequations}
Next, we find at $O(\Phi_w^1\delta^0)$, Eq.~\eqref{Eq: rhoNP} \& \eqref{Eq: poisson} become
\begin{subequations}
\begin{equation}
\begin{array}{c}
    \displaystyle
    \frac{\partial \rho_{10}}{\partial \tau} 
    = \frac{1}{R}\frac{\partial}{\partial R}
    \left(R\frac{\partial}{\partial R}(\rho_{11}+2\Phi_{11})\right)+\frac{\partial^2}{\partial Z^2}\left( \rho_{10}+2 \Phi_{10} \right)
    \label{Eq: NP_order10}
\end{array}
\end{equation}
\begin{equation}
\begin{array}{c}
    \displaystyle
    \frac{1}{R}\frac{\partial}{\partial R}
    \left(R\frac{\partial\Phi_{10}}{\partial R}\right)
    = -\frac{\kappa^2}{2}\rho_{10}.
    \label{Eq: poisson_order10}
\end{array}
\end{equation}\label{Eq:order1}
We note that Eq.~\eqref{Eq: saltNP} is neglected, as it is not required to determine the potential and charge density profiles. We also note that the additional factor of 2 for $\Phi_{11}$ and $\Phi_{10}$ is a result of the leading order salt concentration, $S_{00}=2$. As Eqs.~\eqref{Eq: del_minus_1} and \eqref{Eq:order1} provide the lowest-order non-trivial solution, i.e., $O(\Phi_w^1 \delta^0)$, we limit the analysis to them, as they capture the essential features of charging. However, we would like to eliminate the $\rho_{11} + 2 \Phi_{11}$ from Eq.~\eqref{Eq: NP_order10}.
\par{} To do so, we use the no-flux condition at the wall, $\mathbf{n}\cdot(\boldsymbol{\nabla}\rho+S\boldsymbol{\nabla}\Phi)=0$. Mathematically from Eq.~\eqref{Eq: dimensionalPoreWall}, $\mathbf{n} \propto \nabla \left( r - \alpha^*(z) \right)$, which in dimensionless variables yields
\begin{equation}
    \mathbf{n}\propto\frac{\ell_p^*}{a_p^*}\mathbf{e}_R-\frac{d\alpha}{dZ}\mathbf{e}_Z=\frac{1}{\delta^{1/2}}\mathbf{e}_R-\frac{d\alpha}{dZ}\mathbf{e}_Z.
    \label{Eq: norm}
\end{equation}
Substituting Eq.~\eqref{Eq: norm} in $\mathbf{n}\cdot(\boldsymbol{\nabla}\rho+S\boldsymbol{\nabla}\Phi)=0$ at $R=\alpha(Z)$ and expanding to $O(\Phi_w^1 \delta^0)$, we obtain
\begin{equation}
        \frac{\partial}{\partial R}(\rho_{11}+2\Phi_{11})|_{R=\alpha(Z)}=\frac{d\alpha}{dZ}\frac{\partial}{\partial Z}(\rho_{10}+2\Phi_{10})|_{R=\alpha(Z)}.
        \label{Eq: FluxRelation}
\end{equation} 
Eq.~\eqref{Eq: FluxRelation} imposes an additional constraint on the pore geometry, that being $\alpha(Z)$ must be smooth and $\frac{d\alpha}{dZ}=O(1)$ (note that in the non-stretched coordinates, this would be $O(\delta)$). Sharp variations in wall geometry would violate the expansion at this order. Physically, this requirement can be understood by noting that steep geometric gradients would introduce local flux restrictions within the pore, breaking the assumption of slow axial variation along the pore. Eq.~\eqref{Eq: FluxRelation} can be used to eliminate $\rho_{11} + 2 \Phi_{11}$ in Eq.~\eqref{Eq: NP_order10}. However, since Eq.~\eqref{Eq: FluxRelation} is only valid at $R=\alpha(Z)$, we radially integrate Eq.~\eqref{Eq: NP_order10} to write 
\begin{equation}
\begin{array}{c}
    \displaystyle
    \int_0^{\alpha(Z)} \frac{\partial \rho_{10}}{\partial \tau} R dR
    = \int_0^{\alpha(Z)} \frac{\partial}{\partial R}
    \left(R\frac{\partial}{\partial R}(\rho_{11}+2\Phi_{11})\right) dR + \int_0^{\alpha(Z)} \frac{\partial^2}{\partial Z^2}\left( \rho_{10}+2 \Phi_{10} \right) R dR
    \label{Eq: NP_order10_integrated},
\end{array}
\end{equation}
where the first term on the right hand side can integrated to write 
\begin{equation}
\begin{array}{c}
    \displaystyle
    \int_0^{\alpha(Z)} \frac{\partial \rho_{10}}{\partial \tau} R dR
    = \alpha \left. \frac{\partial}{\partial R}(\rho_{11}+2\Phi_{11})\right|_{R=\alpha(Z)} + \int_0^{\alpha(Z)} \frac{\partial^2}{\partial Z^2}\left( \rho_{10}+2 \Phi_{10} \right) R dR
    \label{Eq: NP_order10_integrated}.
\end{array}
\end{equation}
Finally, substituting the boundary condition from Eq.~\eqref{Eq: FluxRelation} in the above, we obtain the following equation only in $O(\Phi_w^1 \delta^0)$ variables
\begin{equation}
\begin{array}{c}
    \displaystyle
    \int_0^{\alpha(Z)} \frac{\partial \rho_{10}}{\partial \tau} R dR
    = \alpha \frac{d \alpha}{dZ}\frac{\partial }{\partial Z} (\rho_{10} + 2 \Phi_{10})+ \int_0^{\alpha(Z)} \frac{\partial^2}{\partial Z^2}\left( \rho_{10}+2 \Phi_{10} \right) R dR
    \label{Eq: NP_order10_integrated2}.
\end{array}
\end{equation}
Due to Eq.~\eqref{Eq: symmetry}, $ \frac{\partial}{\partial Z} (\rho_{10} + 2 \Phi_{10})$ is constant for all $R$ values, and thus does not need to be specifically evaluated $R=\alpha(Z)$. 
\end{subequations} 

\subsection{Expressing equations in terms of the electrochemical potential}
Hereafter, we omit the subscript notation for $O(\Phi_w^1\delta^0)$ terms. For clarity, we define $\hat{\rho} = \Phi_w\rho_{10}$ and $\hat{\Phi} = \Phi_w \Phi_{10}$. We
also introduce $\hat{\mu}=\Phi_w(\rho_{10}+2\Phi_{10})$, referred to as the electrochemical potential of charge; briefly restoring dimensions, we observe, $\hat{\mu}^* = \frac{k_B T}{e} \hat{\mu}$. This quantity corresponds to the linearized electrochemical potential difference between the cation and the anion, $\hat{\mu}=\mu_+-\mu_-$~\cite{henrique2024network}, and this combined quantity will be shown to be a useful metric for describing the charging dynamics along the pore. Our goal is to express the Nernst-Planck equation at $O(\Phi_w^1\delta^0)$ in terms of $\hat{\mu}$. % Although it is theoretically possible to model pores with steep wall gradients, such configurations are inherently a two-dimensional transient problem that requires significantly greater computational expense and are beyond the scope of this work.
\par We start by noting that Eq.~\eqref{Eq: symmetry} can be transformed into $\hat{\mu}$ to be read as
\begin{subequations}
\begin{equation}
    \frac{\partial\hat{\mu}}{\partial R}=0, \label{Eq: conservationFlux}
\end{equation}
or equivalently, 
\begin{equation}
    \hat{\mu}(Z,\tau)= \hat{\rho}(R,Z,\tau)+2 \hat{\Phi} (R,Z,\tau), \label{Eq: conservationPotential}
\end{equation}
as $\hat{\mu}$ is constant with respect to a given radial cross-section. Eq.~\eqref{Eq: conservationPotential} can also be expressed in terms of the average charge density $(\overline{\rho})$ and potential $(\overline{\Phi})$ within a given radial cross-section,
\begin{equation}
    \hat{\mu}(Z,\tau)=\overline{\rho}(Z,\tau)+2\overline{\Phi}(Z,\tau)\label{Eq: conservationPotential2}.
\end{equation}\label{Eq: conservationEqs}
\end{subequations}
While these quantities are still unknown, $\overline{\rho}$ and $\overline{\Phi}$ only depend on the axial position, as the averages are not functions of the radial coordinate. 
\par{} First, we will employ Eq.~\eqref{Eq: symmetry} and integrate the Poisson equation in Eq.~\eqref{Eq: poisson_order10} to eliminate $\Phi_{10}$ and find the relationship between $\overline{\rho}$ and $\hat{\rho}$. Second, we will radially integrate Eq.~\eqref{Eq: NP_order10} to obtain a self-contained equation in $\hat{\mu}(Z, \tau)$ only.
\par{} Substituting $ - 2\frac{\partial \hat{\Phi}}{\partial R} = \frac{\partial \hat{\rho}}{\partial R}$ from Eq. \eqref{Eq: symmetry} in Eq. \eqref{Eq: poisson_order10} and restoring the variables $\hat{\rho}$, $\hat{\Phi}$ and $\hat{\mu}$, we get 
\begin{subequations}
\begin{equation}
        \frac{1}{R}\frac{\partial}{\partial R} \left( R\frac{\partial\hat{\rho}}{\partial R} \right) =\kappa^2\hat{\rho},
        \label{Eq: poisson_rho}
\end{equation}
which has solutions in terms of modified Bessel functions of order zero,
\begin{equation}
    \hat{\rho}(R,Z,\tau)=C_1(Z,\tau) I_0(\kappa R)+ C_2(Z,\tau) K_0(\kappa R) \label{Eq: besselSoln}. 
\end{equation}
\par{} We note that Eq.~\eqref{Eq: poisson_rho} is implicitly time dependent, since the equation is linear in $\hat{\rho}$ and a separation of variables holds. The time dependency will be introduced by the gradual screening of the double layers along the pore's axis. In order for the solution to remain physical at $R=0$, $C_2=0$. Next, $C_1$ can be written in terms of the average charge density $(\overline{\rho})$ by integrating Eq. \eqref{Eq: besselSoln} from the center to the pore wall
\begin{equation}
     \int_{0}^{\alpha(z)}\hat{\rho} R\,dR = \overline{\rho}\frac{\alpha^2}{2}=\int_{0}^{\alpha(z)}C_1 I_0(\kappa R) R\,dR, 
\end{equation}
or
\begin{equation}
    \hat{\rho}=\overline{\rho}\,\frac{\kappa\alpha}{2} \frac{I_0\left(\kappa R\right)}{I_1\left(\kappa\alpha\right)},
    \label{Eq: rho_rho_bar}
\end{equation}
where $\bar{\rho}(Z,\tau)$ is still unknown. Calculating $\hat{\rho}$ from Eq.~\eqref{Eq: rho_rho_bar} at $R=\alpha$ and utilizing $\hat{\Phi}=\Phi_w$ at $R=\alpha$, Eq.~\eqref{Eq: conservationPotential} yields,
\begin{equation}
\hat{\mu}(Z,\tau)=\overline{\rho}\,\frac{\kappa\alpha}{2}\frac{I_0\left(\kappa\alpha\right)}{I_1\left(\kappa\alpha\right)}+2\Phi_w\label{Eq: mu_wall}.
\end{equation}
\end{subequations}
\begin{subequations}
Next, we substitute $\rho_{10} = \frac{1}{\Phi_w} \hat{\rho}$ and $\left( \rho_{10} + 2 \Phi_{10} \right) = \frac{1}{\Phi_w} \hat{\mu}$ in Eq.~\eqref{Eq: NP_order10_integrated2} to write 
\begin{equation}
    \int_{0}^{\alpha(z)}\frac{\partial\hat{\rho}}{\partial\tau}R\,dR= \alpha \frac{d\alpha}{dz}\frac{\partial \hat{\mu}}{\partial Z}+\int_{0}^{\alpha(z)}\frac{\partial^2\hat{\mu}}{\partial Z^2}R\,dR.
\end{equation}
Using Eq. \eqref{Eq: conservationFlux}, $\hat{\mu}$ is factored out of the integral on the right-hand side of the above equation. Finally, integrating and using $\int_{0}^{\alpha(z)}\hat{\rho} R\,dR = \overline{\rho}\frac{\alpha^2}{2}$ on the left-hand side yields,
\begin{equation}
    \frac{\alpha^2}{2}\frac{\partial\overline{\rho}}{\partial\tau}=\alpha\frac{d\alpha}{dZ} \frac{\partial \hat{\mu}}{\partial Z} +\frac{\alpha^2}{2}\frac{\partial^2\hat{\mu}}{\partial Z^2},
\end{equation}
or equivalently,
\begin{equation}
    \frac{\partial\overline{\rho}}{\partial\tau}=\frac{1}{\alpha^2}\frac{\partial}{\partial Z}\left(\alpha^2\frac{\partial\hat{\mu}}{\partial Z}\right).
\end{equation}
\end{subequations}
Or rewriting $\overline{\rho}$ in terms of $\hat{\mu}$ using Eq.~\eqref{Eq: mu_wall},  we obtain the crucial result of this article, i.e., 
\begin{subequations}    
\begin{equation}
    \alpha^2 \frac{\partial\hat{\mu}}{\partial\tau}=\frac{\kappa \alpha}{2}\frac{I_0(\kappa\alpha)}{I_1(\kappa\alpha)}\frac{\partial}{\partial Z}\left(\alpha^2\frac{\partial\hat{\mu}}{\partial Z}\right)
    \label{Eq: diff_eqn_mu_final}
\end{equation}
\par{}This result showcases a critical feature of the variable pore radius, namely, the transport of the electrochemical potentials of charge by cross-sectional area variations. The term $\frac{\partial}{\partial Z} \left(\alpha^2 \frac{\partial \hat{\mu}}{\partial Z} \right)$ is akin to a diffusion with a variable diffusion coefficient. Upon expanding, Eq. (\ref{Eq: diff_eqn_mu_final}) can be rewritten in the useful form
\begin{equation}
\frac{2}{\kappa \alpha}\frac{I_1(\kappa\alpha)}{I_0(\kappa\alpha)}\frac{\partial\hat{\mu}}{\partial\tau} + U(Z) \dfrac{\partial\hat{\mu}}{\partial Z}=\frac{\partial^2\hat{\mu}}{\partial Z^2},
\end{equation}
\end{subequations} 
where $U(Z) = -\frac{2}{\alpha}\frac{d\alpha}{dZ}$. The impact of change in cross-sectional area is captured by $U(Z) \frac{\partial\mu}{\partial Z}$, which conveys the axial component of wall-normal ion fluxes in the formation of EDLs on a slanted surface {and can be interpreted as a ``pseudo-advective'' term. When $\frac{d\alpha}{dZ}>0$ (a diverging profile), it corresponds to} a negative velocity and transports counter-ions away from the pore end, hindering charging. In contrast, when $\frac{d\alpha}{dZ}<0$, this mechanism accelerates charging. A comparison of the mechanisms of axial diffusion/migration and a tangential component of EDL formation can be estimated by the {magnitude of the ``pseudo-velocity coefficient'' $|\frac{2}{\alpha}\frac{d\alpha}{dZ}|$}.
\subsection{Boundary conditions}
In order to solve Eq. \eqref{Eq: diff_eqn_mu_final}, two boundary conditions---one at the pore entrance and one at the pore end---along with an initial condition are required. The initial condition is straightforward, i.e., at $\tau=0$ there is only an applied potential from the wall and no charge, $\hat{\mu}(Z,0)=2\Phi_w=\mu_w$. The pore end has a no-flux condition, or $\left. \frac{\partial \hat{\mu}}{\partial Z} \right|_{Z=1}=0$. The pore entrance condition requires a more careful analysis since it is connected to reservoir via a SDL. 
\par{} The inclusion of the SDL serves to capture transport limitations through an entrance boundary condition, while also matching the linear potential profiles from DNS. The region outside the pore is modeled as an electroneutral cylinder, in which a flux matching condition is imposed between the SDL and the pore. The geometric length scales of this cylinder are treated as adjustable physical parameters. Although the true ion flux in the external electrolyte is more complex than this representation, this SDL model provides an approximation of the entrance resistance in the linear regime. Inside the SDL, since there are no charged surfaces in the radial direction, the governing equations only vary in the axial direction. Additionally, it is common in continuum simulations to assume that regions outside the pore behave linearly~\cite{gupta2020PRL,janssen2021transmission}. Furthermore, due to small potentials, we can argue that $S_{00}=2$. Therefore, to maintain a constant flux of charge, the electric field must be uniform or equivalently the potential profile linear. As the electric and electrochemical potentials can be related by Eq.~\eqref{Eq: conservationPotential}, the governing equation in this region is
\begin{subequations}
\begin{equation}
    \hat{\mu}= 2 \hat{\Phi} = C_3\cdot Z+C_4.
\end{equation}
In our prior work~\cite{henrique2024network}, we show how the continuity of electrochemical potential of charge is required across an interface. This criterion can be understood from the gradient of the electrochemical potential of charge providing an estimate of charge flux. Across an interface, if there is a difference between electrochemical potentials, there would be a spurious flux of charge, which thus necessitates the electrochemical potential of charge to be continuous. Mathematically, $\hat{\mu}(0^{-}, \tau)=\hat{\mu}(0^{+}, \tau)$  at the pore-SDL interface and $\hat{\mu}(-\ell_s^{*-}, \tau)=\hat{\mu}(-\ell_s^{*+}, \tau)=0$ at the SDL-reservoir interface. Applying these boundary conditions, we write
\begin{equation}
    \hat{\mu}(Z, \tau)=\hat{\mu}(0^+, \tau) \left(1+\frac{Z}{l_s/\ell_p^*} \right),  -\ell_s^*/\ell_p^* \le Z \le 0. 
\end{equation}
Conserving the charge flux across the SDL-pore interface, we write
\begin{equation}   \frac{a_s^{*2}}{\ell_s^*} \left. \frac{\partial \hat{\mu}}{\partial Z} \right|_{Z=0-}= \frac{\alpha^*(0)^{2}}{\ell_p^*}  \left. \frac{\partial \hat{\mu}}{\partial Z} \right|_{Z=0^+}
\end{equation}
The fluxes are governed by the Nernst-Planck equations and can be expressed in terms of the electrochemical potential. Notably, the scaling for the SDL is different than the pore region, requiring a description of the left flux, $(N_\text{left})$, to be expressed in terms of the ratios of length scales,
\begin{equation}
    \left.\text{Bi}\cdot\hat{\mu}\right|_{Z=0^+}=\left. \frac{\partial\hat{\mu}}{\partial Z} \right|_{Z=0^+},\label{Eq: boundaryCondition}
\end{equation}
\end{subequations}

where Bi$=\frac{a_s^{*2}\ell_p^*}{\alpha^{*}(0)^{2} \ell_s^*}$ is the Biot number. 
Examining the limiting behavior of the Biot number offers insight into its physical significance. As $\text{Bi} \rightarrow \infty$, the SDL becomes vanishingly thin, effectively placing the reservoir in direct contact with the pore entrance. In this limit, the SDL imposes no resistance to ion transport from the reservoir into the pore. Conversely, as $\text{Bi} \rightarrow 0$, the SDL becomes infinitely thick, acting as a region of high resistance. In this case, only a negligible flux can pass through the SDL to maintain electroneutrality, effectively isolating the pore from the reservoir and resulting in significantly slower charging dynamics. With the governing equation, boundary conditions, and initial condition defined, the model is complete and can be numerically simulated. Our model's reduced-order enables rapid exploration of different geometries, offering a significant computational reduction over traditional direct numerical simulations.

\subsection{Equivalent circuit model}
 \begin{figure}[t]
    \centering
    \includegraphics[width=0.8\linewidth]{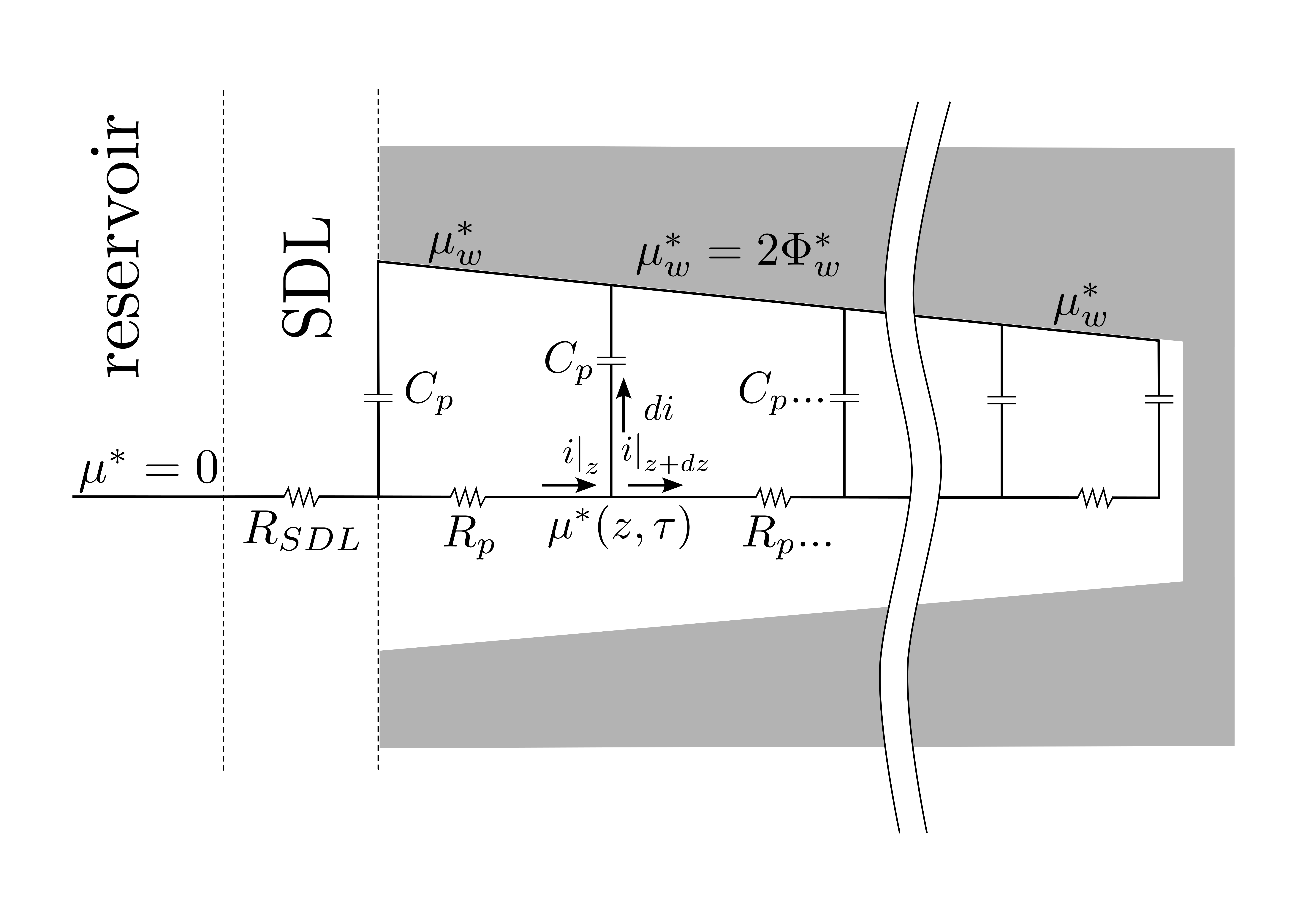}
    \caption{
    A schematic of an equivalent circuit diagram for the dimensional electrochemical potential of charge, $\mu_w^*$, inside a converging pore. The reservoir region acts as the reference potential, with the SDL region bridging the reservoir and the pore regions. This SDL region is treated as a lumped resistor $(R_\mathrm{SDL})$, which is inversely related to the Biot number. Inside the pore, each axial point has an associated resistance $(R_p)$ and capacitance $(C_p)$, which are functions of radius, Debye length, diffusivity, and electric permittivity. Each axial junction has current $i$ flowing through it, and the current $di$ is being stored in the EDL. Driving this electrokinetic flow is an applied wall potential of $\mu_w^*=2\Phi_w^*$.
    }
    \label{fig: equivalentCircuit}
\end{figure}
\par{} Based on Eq.~\eqref{Eq: diff_eqn_mu_final}, we seek to propose an equivalent circuit that incorporates the charging dynamics for an arbitrarily shaped pore; see Fig.~\ref{fig: equivalentCircuit}. This circuit is designed to capture the charging behavior of the electrochemical potential in agreement with the governing equations, which are redimensionalized and rearranged into per-unit-length resistances $(R_p)$ and capacitance $(C_p)$. Our goal is to develop a circuit analogue that aids in the understanding of how pore radius influences charging and how the system behaves in the overlapping and thin EDL limits. Unlike typical circuit models that focus on electrical potential, this formulation centers on the electrochemical potential of charge, $\hat{\mu}$, which is fundamentally distinct. We emphasize that the combined quantity $\hat{\mu} = \hat{\rho} + 2\hat{\Phi}$, and its corresponding gradient, captures the two fundamental driving mechanisms of the system: electromigration and diffusion. By basing our formulation on $\hat{\mu}$, the equivalent circuit remains valid across a broader range of pore sizes and double-layer thicknesses, providing a natural representation of charging behavior from our governing equations.

\par{} In the limit of thin EDLs in a pore, de Levie proposed an RC transmission line model~\cite{de1963porous,de1964porous} that remains widely used today~\cite{movskon2021transmission}. This model yields a Warburg element in an impedance plot, which is commonly argued to be a signature of electrolyte diffusion in a porous material~\cite{cooper2017simulated}. Several modifications to the transmission line model have been proposed, which include nonlinear effects~\cite{biesheuvel2010nonlinear}, Faradaic reactions~\cite{biesheuvel2011diffuse}, and surface conduction~\cite{mirzadeh2014enhanced}. Additionally, these corrections have focused primarily on the thin EDL limit~\cite{cooper2017simulated}. More recently, studies have started to focus on arbitrary Debye lengths~\cite{henrique2022charging, henrique2024network, pedersen2023equivalent, aslyamov2022analytical} and asymmetric diffusivities~\cite{henrique2022impact, henrique2025parallel}. In general, the analyses in the literature focus on straight cylindrical or slit-pores and do not consider a change in pore radius or height. To the best of our knowledge, the only works that consider pore shape and its effect on the equivalent circuit are the works of Black and Andreas~\cite{black2010pore},  Keiser \textit{et al.}~\cite{keiser1976abschatzung} and Cooper \textit{et al.}~\cite{cooper2017simulated}. Black and Andreas~\cite{black2010pore} employ an ad-hoc approach to the equivalent circuit and are not grounded in first principles. While our results bear significant resemblance to those of Keiser \textit{et al.} \cite{keiser1976abschatzung} and Cooper \textit{et al.} \cite{cooper2017simulated}, this agreement holds only in the thin EDL limit. In this regime, the capacitance per unit length scales with the pore's surface area, and the resistance varies inversely with cross-sectional area, consistent with Keiser \textit{et al.}. However, unlike previous models, our approach extends to a broad range of EDL thicknesses, capturing the charging dynamics of both moderate and overlapping double-layers; thus, representing a significant advancement beyond earlier work.

\par{} Starting from Eq.~\eqref{Eq: diff_eqn_mu_final}, we redimensionalize our governing equation:
\begin{equation}    
\alpha^{*2} \frac{\partial \hat{\mu}^*} {\partial t} = \frac{D}{2} \frac{ \left( \alpha^*/\lambda \right) I_0 \left( \alpha^*/\lambda \right) }{ I_1 \left( \alpha^*/\lambda \right) } \frac{\partial}{\partial z} \left( \alpha^{*2} \frac{\partial \hat{\mu}^*}{\partial z} \right),\label{Eq: mu_dimensional}
\end{equation}
where $\hat{\mu}^* = \frac{k_B T}{e} \hat{\mu}$ is the dimensional electrochemical potential of charge, $\lambda$ is the Debye length, $\alpha^*(z)$ is the dimensional radius function, and $I_n$ are modified Bessel functions of the first kind of order $n$. As in de Levie's analysis, we find it convenient to define capacitance per unit length as $C_p$ and the resistance per unit length as $R_p$. Consider a node at an arbitrary axial position $z$ in the circuit of Fig. \ref{fig: equivalentCircuit}. We apply a current balance to this differential node, yielding $di_z=i|_{z}-i|_{z+dz}$. Using Ohm's law, the incoming and outgoing currents into this node can be expressed as,
\begin{subequations}
\label{Eq: eqcircuit_der}
\begin{eqnarray}
\left. i \right|_z = - \left. \frac{1}{R_p}  \frac{\partial \hat{\mu}^*}{\partial z} \right|_{z} , \\
\left. i \right|_{z+dz} = - \frac{1}{R_p}  \left.  \frac{\partial \hat{\mu}^*}{\partial z} \right|_{z+dz}.
\end{eqnarray}
Any current accumulated in this region must be stored as a capacitive current in the EDL, at location $z$. From the definition of capacitance,
\begin{equation}
di_z = \frac{\partial}{\partial t} \left( \hat{\mu}^* - 2\Phi_w^* \right) C_p dz,
\end{equation}
\end{subequations}
where $\Phi_w^*=\frac{k_BT}{e}\Phi_w$ is the applied dimensional wall potential.
\par{} We now seek appropriate expressions for $R_p$ and $C_p$, such that the resulting equivalent circuit formulation is consistent with the governing equation and limiting behavior; hence, before providing the mathematical definitions, we discuss the physical intuition of the result.  Since the total ionic concentration remains constant, the conductivity of the electrolyte also stays constant. Therefore, $R_p$ should be inversely proportional to the local cross-sectional area. In addition, the local capacitance per unit length, $C_p$, should approach different values in the thin and overlapping double-layer limits. In the thin double-layer limit of $\alpha^*/\lambda \rightarrow \infty$, the capacitance per unit surface area should approach the Debye-Huckel limit of $\frac{\epsilon}{\lambda}$, which implies that $C_p = \pi \epsilon \alpha^*/\lambda$ (note that this is a factor of 1/2 different than the de Levie model because $\hat{\mu}^*$ contains a factor of $2\Phi^*$ and hence capacitance has to be adjusted to maintain equal charge). For the limit of $\frac{\alpha^*}{\lambda} \ll 1$, physically, no potential is screened and the charge varies only axially. For $\frac{\alpha^*}{\lambda} \ll 1$, redimensionalizing Eq.~\eqref{Eq: mu_wall}, $\int_0^{\alpha^*} e(c_+ - c_-) 2 \pi r dr = (\hat{\mu}^* - 2 \Phi_w)C_p$ reveals $C_p= \epsilon \pi \frac{\alpha^{*2}}{2 \lambda^2}$. With these limits in mind, we argue that $R_p$ and $C_p$ are given as 
\begin{subequations}
\label{Eq: equivcircuit}
\begin{eqnarray}
R_p=\frac{2 \lambda^2}{D\pi \epsilon \alpha^{*2} }, \\  
C_p(z)=\frac{\pi \epsilon \alpha^*} {\lambda}\frac{I_1(\alpha^*/\lambda)}{I_0(\alpha^*/\lambda)}.\label{Eq: capacitance}
\end{eqnarray} 
\end{subequations}
Substituting these definitions for resistance and capacitance into the current balance yields,
\begin{equation}
    R_pC_p\frac{\partial\mu^*}{\partial t}=\frac{2 \lambda^2}{D \pi \epsilon \alpha^{*2} }\frac{\pi \epsilon \alpha^*} {\lambda}\frac{I_1(\alpha^*/\lambda)}{I_0(\alpha^*/\lambda)}\frac{\partial\mu^*}{\partial t}=\frac{\partial}{\partial z} \left( \alpha^{*2}\frac{\partial\mu^*}{\partial z}\right)
\end{equation}
which is identical to Eq.~\eqref{Eq: mu_dimensional}. Finally, we note that resistance per unit length in the SDL, $R_{p, SDL}$, will simply be scaled by the area of the SDL, or  $R_{p, SDL} = \frac{2 \lambda^2}{\pi D \epsilon a_s^{*2}  }$. While we propose this interpretation of the equivalent circuit; alternative formulations are possible. The present representation should be viewed primarily as a conceptual tool to interpret the charging dynamics from a transmission-line perspective, rather than as a unique or definitive model.
\subsection{Pore descriptions}
We focus on the transient charging behavior inside conical and cylindrical geometries, although our framework can be readily incorporated into other geometries. To describe the conical geometries, we employ,
\begin{equation}
\alpha(Z) = b +  m Z
\end{equation}
where $m<0$ corresponds to a converging pore and $m>0$ to a diverging pore. For the converging geometry, $b=2$ and $m=-1$, whereas for the diverging geometry, we used $b=1$ and $m=1$. The values of $b$ and $m$ are chosen such that the converging and diverging pores represent the same geometry but with the entrance and ends flipped. Since there are no reactions and $\hat{\mu}\to0$ at equilibrium, the charge density and potential profiles of both the converging and diverging pores are identical, but mirrored. For the wide pore, we choose $b=2$ and $m=0$, and for the narrow geometry, $b=1$ and $m=0$. The non-dimensional geometries remain fixed throughout the paper, and all discussion refers to these geometries, unless otherwise noted. To elucidate geometric effects on the transient charging behavior, we analyze charge density distributions, compare electromigrative and diffusive fluxes, and vary entrance resistance. For model validation, we compare our reduced-order framework to a fully resolved Planck–Nernst–Poisson simulation, demonstrating excellent agreement to DNS while providing major computational savings. It is important to note that $\kappa$, or the ratio of reference length scale $(a_p^*)$ with the Debye length $(\lambda)$, is held constant for all the simulations at $\kappa=2$, unless otherwise noted. While some quantitative values change, we show in the supplementary information that the trends hold for $\kappa=0.1$ and $\kappa=10$ as well; see Fig.~\ref{fig:largekappa}\&\ref{fig:smallkappa}
\par{}After solving Eq. \eqref{Eq: diff_eqn_mu_final} for the four geometries, we obtain $\hat{\mu}(Z, \tau)$, which we substitute in Eq. \eqref{Eq: rho_phi_from_mu} to obtain $\hat{\rho}(R,Z,\tau)$ and $\hat{\Phi}(R, Z,\tau)$. The calculations were performed for $\Phi_w=0.5$ for all simulations, except the comparison to the direct numerical simulation (DNS), where at 10mV, $\Phi_w\approx0.389$. At $\tau=0$, the pore is uncharged but set to be at a potential of $\Phi_w$. Hence, $\hat{\mu}(Z,0) = 2 \Phi_w = \mu_w$ for the current discussion. A value of $\hat{\mu} = \mu_w$ corresponds to an uncharged region of the pore, whereas values approaching $\hat{\mu} \to 0$ indicate a charged region. At equilibrium, $\hat{\mu}(Z,\tau) = 0$. Although the time is nondimensionalized by the pore diffusion time, $\tau=Dt/\ell_p^{*2}$, this scaling does not reflect the pore charging timescale as other factors affect charging times more: pore geometry, entrance resistance, and the Debye length ratio $\kappa$. For instance, in the thin double-layer limit, the formation of the EDL occurs on a much shorter timescale; these trends are illustrated in the $\kappa = 10$ case found in Fig.~\ref{fig:largekappa}.
\par{} For a fair comparison, we kept the length scales of the SDL constant across all simulations, except the DNS comparison, and appropriately adjusted Biot number based on the definition of $\text{Bi}$. We assume that $\ell_s^*=\ell_p^*$ and $a_s^*=4a_p^*$. As a result, the Biot numbers are set at $\text{Bi}=4$ for the wide entrances and $\text{Bi}=16$ for narrow ones. Although the SDL is modeled as a cylinder with two characteristic length scales, other constructions are possible. This SDL formulation provides a controlled way to represent entrance resistance, while still preserving the effect of entrance size on the charging dynamics. By fixing the length scales in the SDL region, the size of the transport environment outside the pore remains consistent across cases with different entrance radii. Other constructions---such as those that hold the entrance resistance (or Biot number) constant---do not necessarily share this feature. For example, if the Biot number is fixed, the resulting entrance boundary conditions become identical for both wide and narrow entrances, making a fair comparison difficult.

\subsection{Direct numerical simulation comparison}

\begin{figure}[ht!]
    \centering
    \includegraphics[width=1\linewidth]{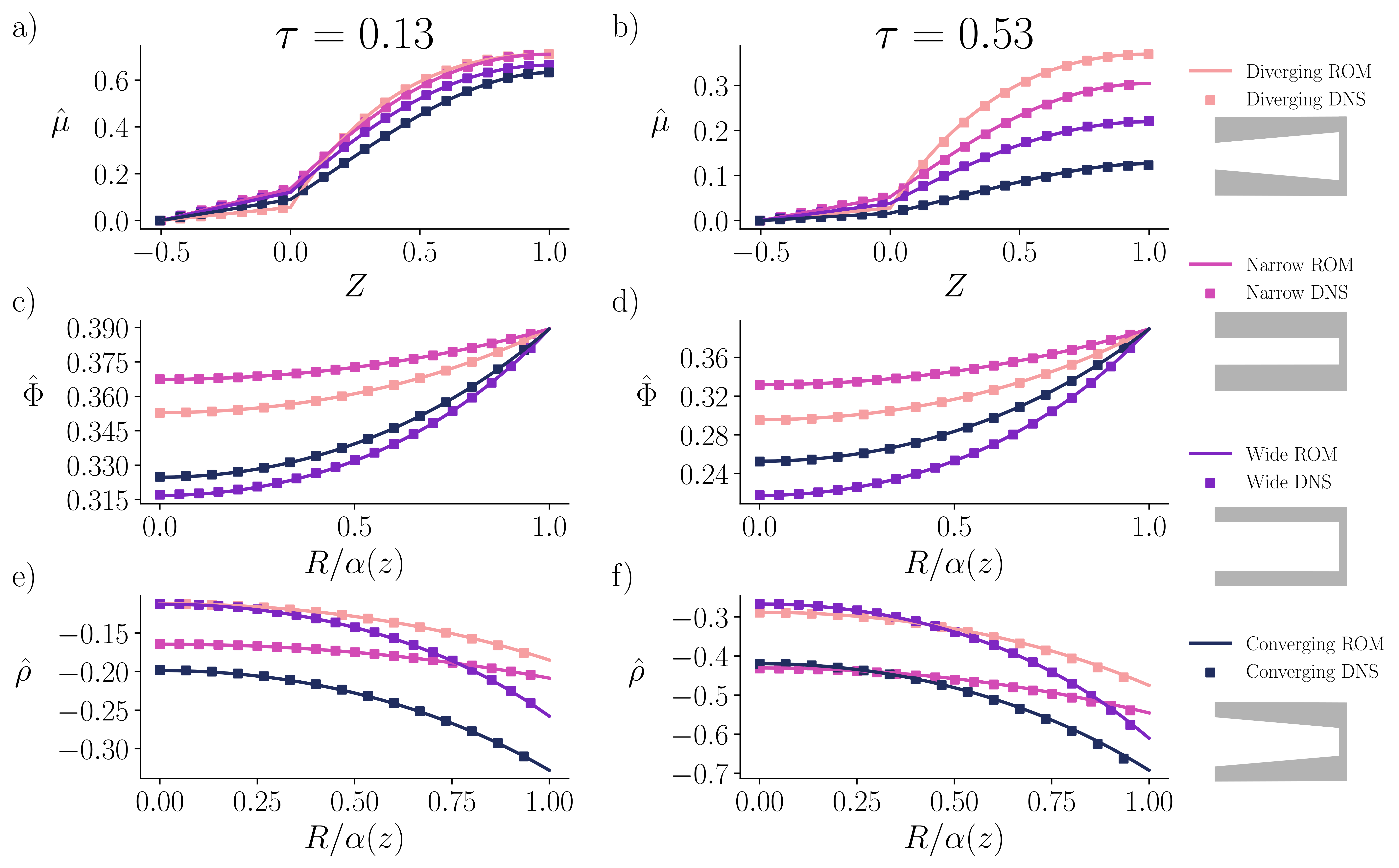}
    \caption{
    The electrochemical potential $\hat{\mu}(Z,\tau)$, the electric potential $\hat{\Phi}(Z,R,\tau)$, and the charge density $\hat{\rho}(Z,R,\tau)$ plotted at various times---early (a,c,e) and intermediate times (b,d,f)---against direct numerical simulations. (a-b) represent plots of the electrochemical potential as a function of axial position, and include the SDL region when $Z<0$. The radial electric potential (c-d) and charge density (e-f) profiles are plotted midway, $(Z=0.5)$, through the pore as a function of radial position. Schematics of all the pores are depicted on the right side of the figure: diverging, narrow, wide, and converging. The entrance-to-end ratio for the converging and diverging cases are 2:1 and 1:2, respectively, with corresponding equations $\alpha(Z) = 2-Z$ and $\alpha(Z) = 1+Z $, respectively. The wide $\alpha(Z)=2$ and narrow $\alpha(Z)=1$ pores represent the widest and narrowest radii for the changing radius cases. $\tau$ denotes the non-dimensional time, $Z$ the non-dimensional length along the pores, and $R$ the non-dimensional radial length.
    }
    \label{fig:mu_plots_conical}
\end{figure}

\par
\begin{figure}[ht!]
    \centering
    \includegraphics[width=1\linewidth]{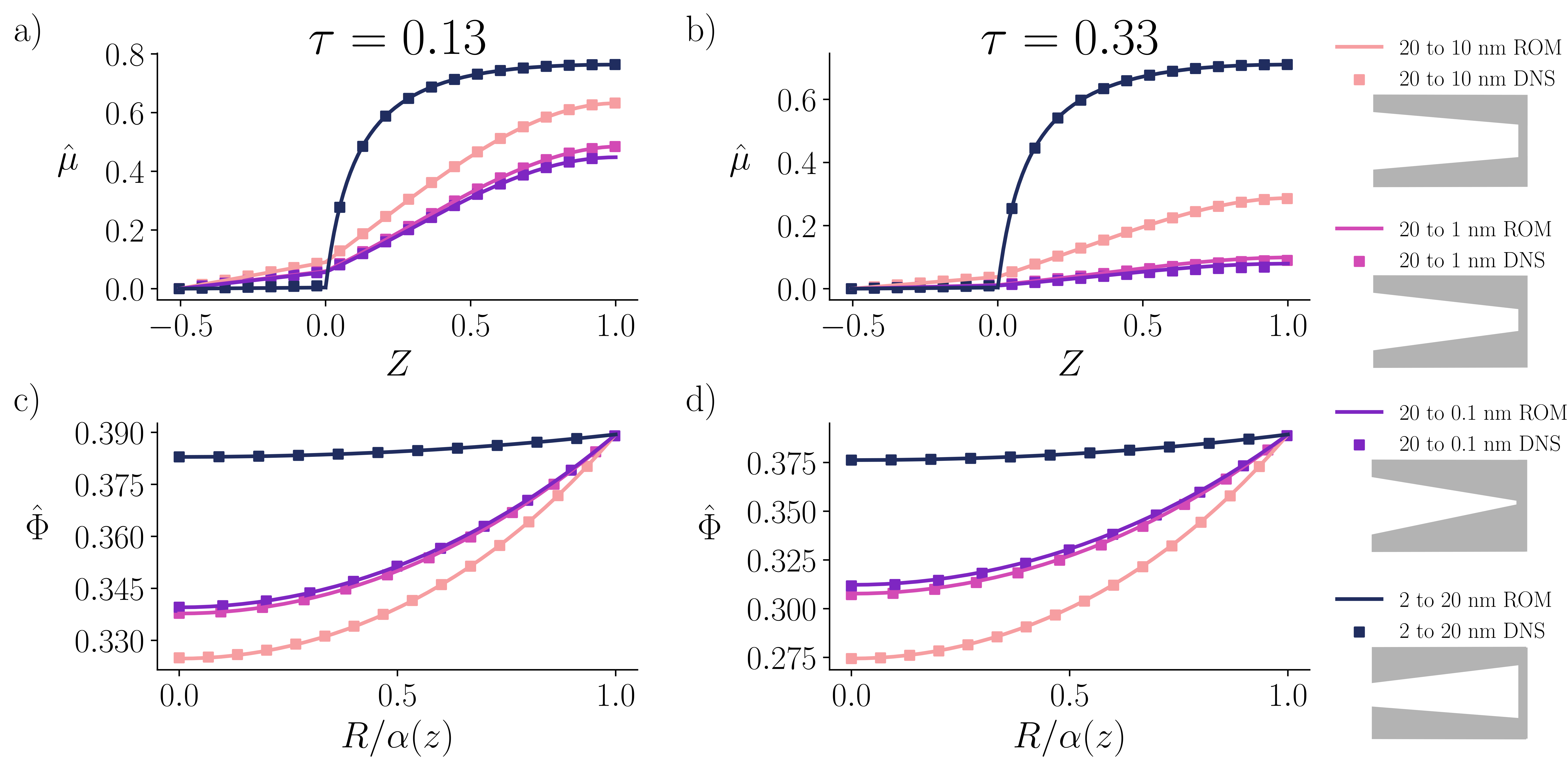}
    \caption{
    {The electrochemical potential $\hat{\mu}(Z,\tau)$ and the electric potential $\hat{\Phi}(Z,R,\tau)$ plotted at various times---early (a,c) and intermediate times (b,d)---against direct numerical simulations with larger changes in slope. (a-b) represent plots of the electrochemical potential as a function of axial position, and include the SDL region when $Z<0$. The radial electric potential (c-d) profiles are plotted midway, $(Z=0.5)$, as a function of radial position. Three converging pores and one diverging pore are tested. The converging pores share the same entrance radius of 20 nm, but differing in their end radii. The 20–10 nm case corresponds to the converging pore discussed previously, while the 20–1 nm and 20–0.1 nm pores explore steeper slopes, testing the limits of the perturbation analysis. Finally, the diverging pore is also added with a radius starting of 2 nm and an ending radius of 20 nm.}
    }
    \label{fig: LargeSlopes}
\end{figure}

{} To validate the reduced-order model (ROM), we compared its predictions directly against DNS of the full nonlinear PNP equations. We find close agreement between the DNS and ROM, demonstrating that the reduced model captures the influence of geometry on the essential features of pore charging, like the potential and charge distribution profiles. Moreover, additional DNS comparisons, Fig.~\ref{fig: LargeSlopes},  demonstrate that the ROM retains accuracy even for larger wall slopes. For simulation parameters, we set the lengths of the cylindrical SDL to be, $\ell_s^*=0.5\ell_p^*+3\text{max}(\alpha(Z)^*/2)$, and the radii to be $a_s^*=2\text{max}(\alpha(Z)^*)$. Therefore the resulting Biot numbers are $\text{Bi}\approx31.81$ for the diverging pore and $\text{Bi}\approx7.95$ for all other pores, in Fig.~\ref{fig:mu_plots_conical}. The choice of dimensionless group in the DNS was largely driven by numerical convenience. In dimensional terms, the simulated pores are $1 \ \mu\text{m}$ in length and have varying radii ranging from $20 \ \text{nm}$ down to $0.1 \ \text{nm}$, with a characteristic Debye length fixed at $10 \ \text{nm}$. As we show later, we can recover the spatiotemporal variations of charge density, potential, and the electrochemical potential, giving us confidence to investigate the trends from our reduced-order model systematically.
\par{} To compare our linearized reduced-order model against DNS, we plot $\hat{\mu}(Z, \tau)$, $\hat{\Phi}(Z=0.5, R, \tau)$, and $\hat{\rho}(Z=0.5, R, \tau)$ for the four different geometries at two different times. First, we note that we obtain excellent agreement between DNS and our proposed model, both for axial variation as well as the radial variation. This demonstrates that our model can capture the physics of EDL charging in the small potential limit for shape-changing pores without any loss of accuracy.  \par{} The early time dynamics are represented by $\tau=0.13$; see Fig.~\ref{fig:mu_plots_conical}(a,c,e). One would expect that the dynamics of the converging and diverging pores would lie between the dynamics of the narrow and wide pores. However, the diverging pore with $\alpha(Z)=1+Z$ charges slower throughout the majority of the pore compared to all other cases, and the converging pore $\alpha(Z)=2-Z$ charges faster than all other cases. Although at early times, the diverging pore still maintains a slightly more developed EDL near the entrance, compared to the converging pore. 
\par{}The majority of this entrance behavior at early times can be attributed to the pore mouth size, consequently affecting the Biot number. The fixed length scales of the SDL region result in a higher Biot number for narrow entrances. This enhances the entrance flux, leading to more rapid charging near the pore mouth, as observed in the diverging case. A high Biot number effectively reduces entrance resistance, enabling greater electrochemical flux into the system. Despite the higher entrance flux observed in the diverging case, the overall charging rate across the entire pore is lower---a result influenced by several factors that will be discussed later.
\par{}At later times, as shown in Fig.~\ref{fig:mu_plots_conical}(b,d,f), the trends observed at early times persist and become more substantial. At intermediate times, the difference in charging rates between the diverging–narrow pores and the converging–wide pores becomes more pronounced. The convergent pore exhibits more well-formed electric double-layers compared to the other cases, even near the entrance. Additionally, we observe that the diverging pore shows a greater separation in the electrochemical potential profile than the narrow pore.
\par{}The change in charging rate for later time is likely due to a multitude of factors, but we will highlight a few mechanisms here. From Eq.~\eqref{Eq: diff_eqn_mu_final}, it is evident that the cross-sectional area plays a significant role in determining the electrochemical behavior within the pore. In the converging geometry, the cross-sectional area decreases along the pore length, while in the diverging case, it increases. Consider two connected differential slices within the pore. Any current not accumulated by the EDL would be affected by the locally changing cross-sectional area, altering the electrochemical flux. Specifically, if the adjacent region has a smaller area, the electrochemical flux must increase to maintain the current, as seen in the converging pore. This results in higher local fluxes and consequently, more rapid charging. Conversely, in the diverging pore, the increasing cross-sectional area causes a reduction in local flux, slowing the charging process. Furthermore, as the pore widens or narrows, the total double-layer capacitance changes accordingly, increasing with a wider cross-section and decreasing with a narrower one.
\par{}Fig.~\ref{fig:mu_plots_conical}(c-d) shows the simulated radial potential profiles at $Z=0.5$ at early and intermediate times. At this location, the converging and diverging pores share the same radius, and at equilibrium the potential and charge density profiles would be identical. For convenience and readability, the radial coordinate is normalized by the local wall distance, $\alpha(Z)$. At early times, we observe that the potential profiles are relatively undeveloped, remaining close to the applied potential. Additionally, all the potential profiles monotonically increase from the pore center towards the wall, reaching the applied wall potential $\Phi_w\approx0.389$. The largest potential difference from the centerline to the wall occurs for the wide and converging geometries, consistent with their faster charging dynamics. In contrast, the diverging and narrow pores exhibit a smaller potential difference, indicative of slower charging. At intermediate times, these trends persist. Although the relative difference between individual profiles decreases, the absolute potential drop from the center to the wall continues to grow as the EDL becomes more developed.
\par{}Similar to the potential profiles, the corresponding charge density distributions are shown in Fig.~\ref{fig:mu_plots_conical}(e-f). These profiles provide insight into the structure and charge state of the local EDL, as the charge density is directly related to it. In early times, we observe that the converging pore exhibits the highest charge density, then followed by the narrow pore. In contrast, the diverging and wide pores show lower charge densities through the radial cross section. Notably, the wide pore displays a more developed EDL near the wall than either the narrow or diverging pore. The narrow and diverging pores have relatively flat charge density profiles, while the wide and converging geometries display steeper, monotonically decreasing profiles. This behavior is indicative of more developed double-layers near the wall---which follows from their electrochemical potential profiles depicted in Fig.~\ref{fig:mu_plots_conical}(a-b)---suggesting that these pores are closer to equilibrium conditions. These trends persist at intermediate times, with the profiles becoming more pronounced and steeper, similar to the potential profiles.
\par{}A key strength of our reduced-order model is its accurate descriptions of the potential and charge density profiles. Fig.~\ref{fig:mu_plots_conical}(c-d) shows that the predicted potential profiles, given by simulating the full PNP equations, match very closely to those given by perturbation analysis with reduced computational cost. Similarly, Fig.~\ref{fig:mu_plots_conical}(e-f) demonstrates that the charge density recovered from the electrochemical potential agrees with directly simulating both ions using the PNP equations. Additionally, the constructed quantity $\hat{\mu}$---the electrochemical potential of charge---which is not explicitly included in the DNS, can be computed using Eqs.~\eqref{Eq: conservationEqs}. As shown in Fig.~\ref{fig:mu_plots_conical}(a-b), $\hat{\mu}$ demonstrates its effectiveness in characterizing the charging behavior, exhibiting strong agreement with the converted DNS data, and providing a useful charging metric.

\par{} To further validate our ROM and probe its limitations, we performed DNS of the nonlinear PNP equations for modified converging and diverging pores with larger changes in radii; see Fig.~\ref{fig: LargeSlopes}. Due to numerical limitations, only one diverging pore geometry was tested. Simulation parameters were kept consistent with the other DNS cases, allowing for a systematic comparison with ROM predictions. Three converging pores were tested, all with entrance radii fixed at $20 \text{ nm}$ and end radii of $10 \text{ nm}$, $1 \text{ nm}$, and $0.1 \text{ nm}$, representing $2\times$, $20\times$, and $200\times$ decreases in radii. The $20 \rightarrow 10 \text{ nm}$ case corresponds to the original converging pore discussed previously. Because all converging pores share the same entrance size, they also share the same boundary condition, with $\text{Bi}\approx7.95$. The single diverging case tested had an entrance radius of $2 \text{ nm}$ and a final radius of $20 \text{ nm}$; because its entrance size differs, the corresponding Biot number is $\text{Bi}\approx795$.
\par{} Despite these strong geometric variations, the ROM remains in close agreement with DNS, accurately capturing the evolution of both the electrochemical potential $\hat{\mu}(Z,\tau)$ and the electric potential $\hat{\Phi}(Z,R,\tau)$ across the pore. The ROM slightly overpredicts the charge state, with DNS showing the pores closer to equilibrium than the ROM predicts for the converging radii cases $20 \rightarrow 1 \text{ nm}$ and $20 \rightarrow 0.1 \text{ nm}$. Additionally, the acceleration in the charging rate exhibits diminishing returns: the increase from $20 \rightarrow 10 \text{ nm}$ to $20 \rightarrow 1 \text{ nm}$ is larger than the subsequent increase from $20 \rightarrow 1 \text{ nm}$ to $20 \rightarrow 0.1 \text{ nm}$, indicating a regime in which further geometric constrictions produce only marginal improvements. The steeper diverging case ($2 \rightarrow 20\,\text{nm}$) shows additional charging rate slowdown compared to the $10 \rightarrow 20\,\text{nm}$ diverging case (Fig.~\ref{fig:mu_plots_conical}) at the same non-dimensionalized time. Additionally, the steeper diverging case exhibits a greater separation in charging rate relative to the original converging case, indicating that these widening pore openings substantially reduce the charging rate.
\par{} Overall, the DNS confirms that the ROM accurately captures the essential physics of pore charging and demonstrates that the electrochemical potential is a useful metric for quantifying charging dynamics in geometrically varying pores. While minor discrepancies exist between the ROM and DNS, particularly at later times, the close agreement at early times reinforces the model's validity. Even with greater variations in radii, the ROM provides reliable qualitative insight into the influence of pore geometry on charging dynamics.

\subsection{Geometry and charging}
\begin{figure}[ht!]
    \centering
    \includegraphics[width=1\linewidth]{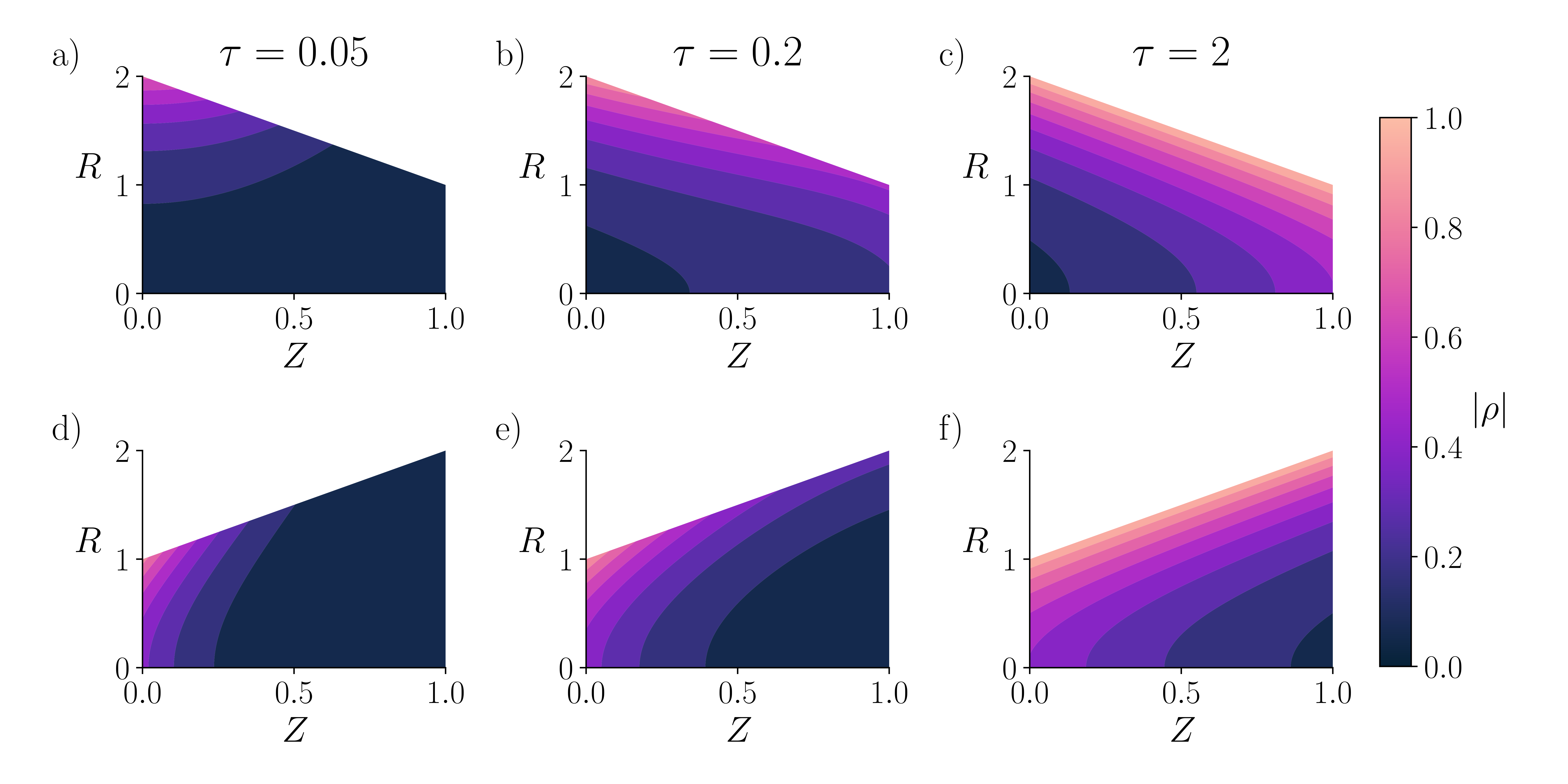}
    \caption{
    Contour plots of charge density ($\hat{\rho}$) throughout the pore at various times. (a-c) depict the converging geometry, and (d-f) depict the diverging geometry, described by $\alpha(Z)=2-Z$ and $\alpha(Z) = 1+Z$, respectively. The charge density contours are shown at early (a,d), intermediate (b,e), and equilibrium (c,f) times. Lighter regions correspond to higher charge densities, while darker regions represent a more neutral state. $\tau$ denotes the non-dimensional time, $Z$ the non-dimensional length along the pores, and $R$ the non-dimensional radius. The initial and system conditions are $\hat{\mu}(Z,\tau=0) = 1$, $\kappa=2$, $\ell_s^*=\ell_p^*$ and $a_s^*=4a_p^*$.
    }
    \label{fig:edlContour}
\end{figure}
\par{} The formation of the EDL for the convergent and divergent geometries are depicted in Fig.~\ref{fig:edlContour}, where brighter contours indicate areas of high charge density. Fig.~\ref{fig:edlContour}(a,d), which represent the early charging dynamics, show that the EDL penetrates deeper into the divergent pore compared to the convergent pore. For the intermediate charging dynamics, shown in Fig.~\ref{fig:edlContour}(b,e), the EDL is more fully developed, as indicated by thicker contour sections in the convergent plot. The convergent pore exhibits a higher charge state with its thicker EDL. The equilibrium solutions are shown by Fig.~\ref{fig:edlContour}(c,f). At this stage, the charge distributions are mirror images of each other, as the convergent and divergent pores are geometric reflections of each other. At equilibrium $\hat{\mu} \rightarrow 0$, resulting in time-independent charge and potential profiles from Eqs.~\eqref{Eq: conservationEqs},
\begin{subequations}
    \begin{align}
        \rho &= - 2\Phi_w \frac{I_0(\kappa R)}{I_0(\kappa\alpha)} \label{Eq: rho_ss} \\
        \Phi &= \Phi_w\frac{I_0(\kappa R)}{I_0(\kappa\alpha)} \label{Eq: phi_ss}
    \end{align}
\end{subequations}
where $\alpha$ is a function axial position. Additionally, see supplemental video 2 for visualization of charge density contours over time.
\par{}

\begin{figure}[h]
    \centering
    \includegraphics[width=1\linewidth]{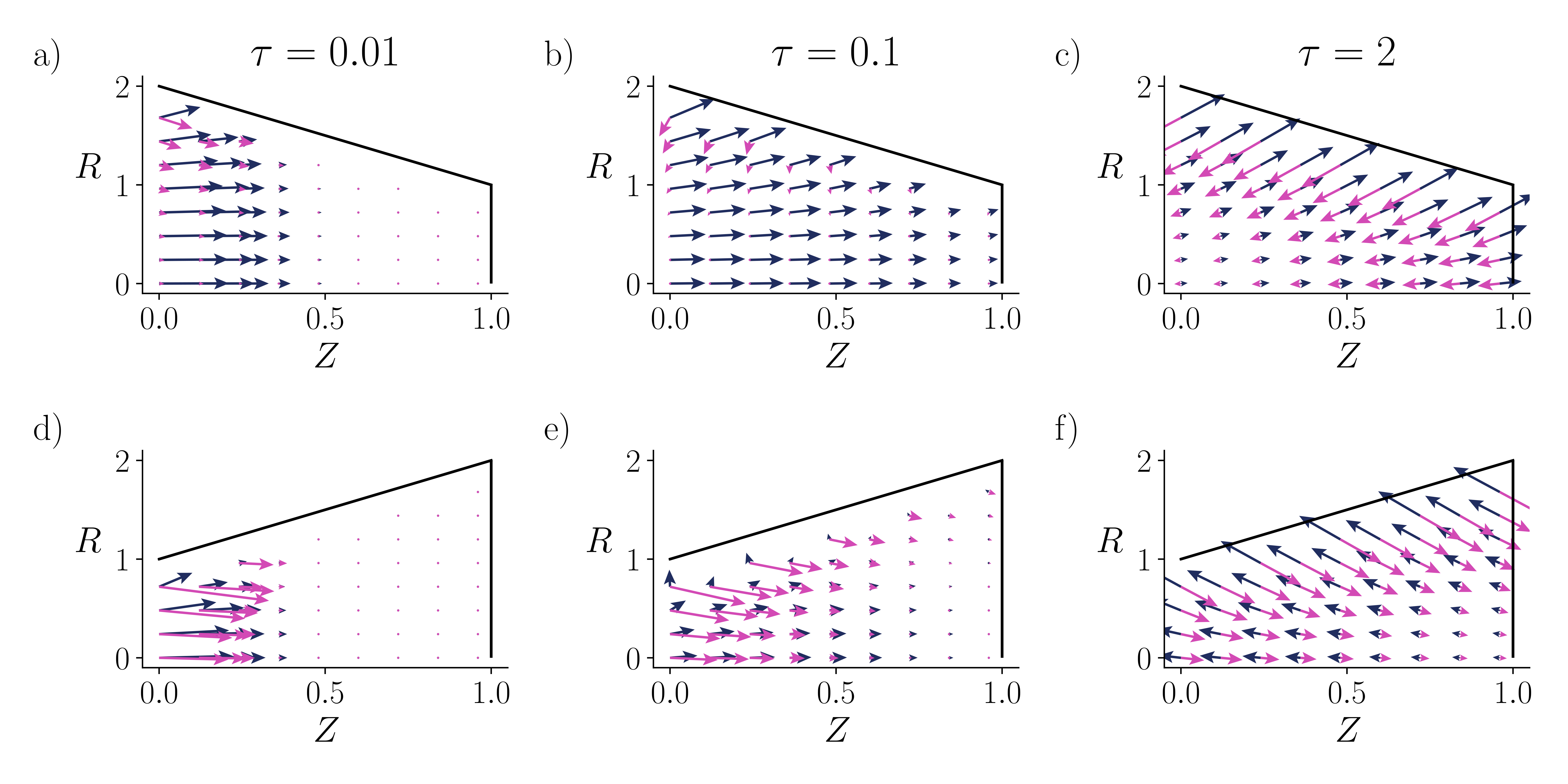}
    \caption{
    Quiver plot of $\boldsymbol{\nabla}\hat{\rho}(Z, R,\tau)$ and $\boldsymbol{\nabla}\hat{\Phi}(Z, R,\tau)$, representing the diffusive (light pink) and electromigrative (dark blue) fluxes, respectively. The flux profiles within the pores at early (a,d), intermediate (b,e), and equilibrium (c,f) times. The entrance-to-end ratio set at 2:1 for the converging pore and 1:2 for the diverging pore, defined by $\alpha(Z) = 2-Z$ and $\alpha(Z) = 1+Z$, respectively. $\tau$ denotes the non-dimensional time, $Z$ the non-dimensional axial coordinate, and $R$ the non-dimensional radial coordinate. The initial and system conditions are $\hat{\mu}(Z,\tau=0) = 1$, $\kappa=2$, $\ell_s^*=\ell_p^*$ and $a_s^*=4a_p^*$.
    }
    \label{fig:flux}
\end{figure}

\par{} A useful and intuitive way to think about this system is via the two different driving forces and their respective fluxes: the electromigrative (dark blue) and diffusive (light pink) fluxes. Fig.~\ref{fig:flux} illustrates a quiver plot of these two driving fluxes for $\kappa=2$ at three representative time points: early, intermediate, and equilibrium. At the entrance, we find that electromigrative fluxes dominate over diffusion for the converging pore, whereas the electromigrative fluxes and diffusion are comparable for the diverging pore. This is expected because a smaller $\kappa \alpha$ implies a diffusion-dominated charging, whereas a larger $\kappa \alpha$ implies that electromigration becomes the dominant transport mechanism~\cite{henrique2022charging}. Therefore, for $\kappa=0.1$, both pores are dominated by diffusion, whereas for $\kappa=10$, both pores are dominated by electromigration; see Supplementary Information. We now emphasize why the charging of EDL speeds up for all $\kappa$ values. 
\par{} The electromigrative flux serves two purposes. First, it pushes the counter-ions towards the walls and the co-ions away from the walls, thus assisting with the formation of EDL. Second, electromigration also charges the pore axially, along with some help from diffusion. Since the normal vector to the wall points towards the end of the pore for the converging scenario (Fig.~\ref{fig:flux} (a-c)),  electromigration retains a favorable axial component, and the two purposes of electromigration assist each other. In contrast, for the diverging pore (Fig.~\ref{fig:flux} (d-f)), the axial component of the electromigrative flux switches the direction, and the two purposes compete, slowing down the charging process. A similar trend is readily observable for $\kappa=10$; see Supplementary video 1 and figures.  
\par{} The discussion above highlights how a converging pore accelerates charging by using favorable electromigrative fluxes. However, for small $\kappa$ values, electromigration ceases to exist; see Supplementary Fig.~\ref{fig:smallkappa}\&\ref{fig:largekappa}. Even then, the converging pore charges faster. While the changing cross-sectional area and favorable electromigrative directions may initially appear to be two different effects, they are identical phenomena, since a varying cross-section necessarily affects the tangential wall component. Notably, the charging rate increases in converging geometries becomes more pronounced as the Debye length decreases.
\par
\begin{figure}[h!]
    \centering
    \includegraphics[width=1\linewidth]{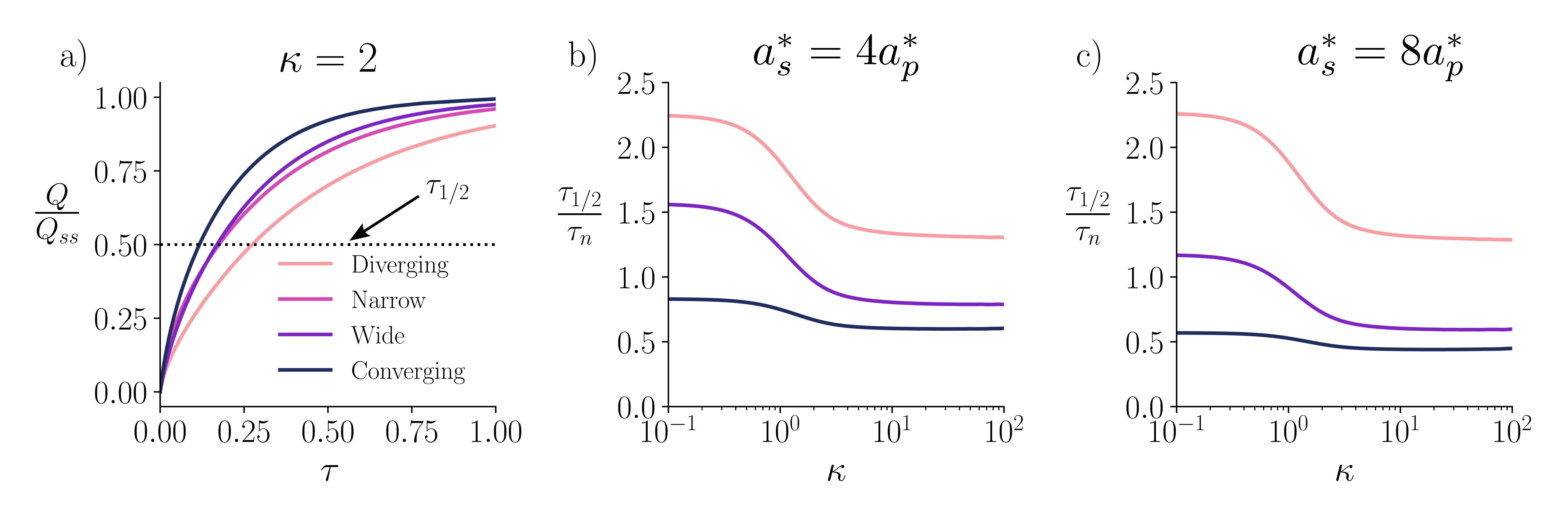}
    \caption{
    Plot of charge fraction---quantified as total instantaneous charge $(Q)$ over total steady state charge $(Q_{SS})$---as a function of time (a), where $\hat{\mu}(Z,\tau=0) = 1$, $\kappa=2$, and $a_s^*=4 a_p^*$. The dotted line represents the $\tau_{1/2}$ or the time it takes a given pore to reach half of its equilibrium charge. For comparison, in plots (b-c), each $\tau_{1/2}$ is normalized by the half charge time of the narrow pore $\tau_n$. This ratio plotted on logarithmic scale versus $\kappa$ for two different characteristic length scales of the SDL. These plots demonstrate the effect of entrance resistance; specifically, plot (b) exhibits a higher entrance resistance than plot (c).
    }
    \label{fig:ChargeVsTime}
\end{figure}

{} We seek to quantify how the Debye length and entrance resistance interplays with charging rate; Fig.~\ref{fig:ChargeVsTime} demonstrates the influence of these two mechanism. Although the electrochemical potential of charge describes the "distance" from equilibrium, it is not analogous to total charge storage within the pore. Since the capacitance changes as a function of radius---larger regions having higher capacitance and narrower regions with lower capacitance---the amount and distribution of charge are important considerations for the transient dynamics. For this, we introduce two new metrics $Q(\tau)$, the instantaneous total charge, and $Q_{ss}$, the equilibrium charge of the pore. Their ratio, $Q/Q_{ss}$, serves as a metric for a pore's charge fraction. This charging metric provides a more comprehensive view of charging, as it represents a weighted average over all sections of the pore. These quantities can be simply derived from taking a volume integral of the average charge density at different points in time, 
\begin{equation}
    Q=\int\limits_V\hat{\rho}(Z,R,\tau)dV=\int\limits_V\bar{\rho}(Z,\tau)dV.
\end{equation}
\par{}From Fig.~\ref{fig:ChargeVsTime}(a), the general trends in charging rates---converging, wide, narrow, and then diverging---continue to hold over time. Consistent with earlier observations, the narrow pore charges more rapidly at early times, but the wide pore eventually outpaces the narrow pore. The higher Biot number initially provides a boost to the narrow pore, but the wide pore's lower resistance allows for faster charging at late times. Interestingly, their overall charging rates, at $\kappa=2$, are relatively similar, unlike those of the converging and diverging pores. The converging pore demonstrates the most substantial improvements in charging rate compared to all other pore cases. In contrast, the diverging exhibits a significant decrease in charging rate. The charging rate decrease is even prominent at early times, where the higher Biot number benefits the narrow pore but not the diverging pore. This is because the total charge fraction includes contributions in the latter sections of the pore, where the narrow pore provides a constant resistance and capacitance throughout. In contrast, the diverging pore struggles to charge these regions throughout time. The charging behavior is marked by a rapid initial charging phase, followed by a slow progression to equilibrium.
\par{}Important considerations are the effects of the Biot number and Debye length on charging dynamics. In Fig.~\ref{fig:ChargeVsTime}(b-c), we compare charging rates with different length scales of the SDL, varying the relative dimensionless Debye ratio, $\kappa$, on a logarithmic scale. The SDL framework is a well-established technique for modeling the transition from reservoir to confined regions~\cite{henrique2022charging,biesheuvel2010nonlinear,biesheuvel2011diffuse}. Changing the characteristic length scale of the SDL affects the Biot number, which quantifies the resistance of electrochemical flux from the bulk region into the pore. We chose to change the SDL radial length scale, $a_s^*$, which is related to $\text{Bi} \propto a_s^{*2}$ from Eq.~\ref{Eq: boundaryCondition}. In the limit $\text{Bi}\rightarrow\infty$, entrance resistance vanishes, whereas $\text{Bi}\rightarrow0$ corresponds to an infinitely resistive pore entrance. The influence of Biot number on the charging dynamics is nuanced, as it has a complex interplay with geometry and Debye length. 
\par{}To evaluate charging rates, we introduce a new metric: the half-charge time, $\tau_{1/2}$, defined as the time required for a pore to reach 50\% of its steady state charge. For clearer visualization and relative comparison, we normalize each pore's $\tau_{1/2}$ by that of the narrow pore $\tau_n$. In Fig.~\ref{fig:ChargeVsTime}(b-c), we note two key limiting behaviors: the overlapping double-layer regime at low $\kappa$, and the thin double-layer regime at high $\kappa$, with a transition occurring around $\kappa=O(1)$. We observe that high $\kappa$ increase the charging rate generally, but the converging pore charges faster---compared to the narrow pore. Additionally, this effect is more prominent with lower entrance resistance, or equivalently, greater $a_s^*$. This behavior stems from the reduced entrance resistance at high Biot, which permits greater ion flux into the pore---benefiting geometries where transport is more limited by entrance effects, like the wide and converging pores. In contrast, diverging pores show minimal improvement, as their performance is primarily constrained by geometric inefficiencies rather than entrance resistance.
\par{}A surprising result emerges in the overlapping double-layer regime, where we found wide pores performing worse than narrow ones. While many studies highlight that wide pores tend to charge faster than narrow ones~\cite{de1963porous,henrique2022charging,aslyamov2022analytical,janssen2021transmission,yang2022simulating}, Fig.~\ref{fig:ChargeVsTime}(b,c) shows that entrance effects are an important consideration. While prior MD studies have identified non-monotonic charging behavior in nanopores~\cite{mo2020ion}, our work highlights a distinct continuum based mechanism of a similar phenomena. We attribute the reversal of the wide-pore charging trend primarily to the Biot number: by limiting the current entering the pore, the higher capacitance of the wider pore slows its charging rate. This effect manifests in the transition region from the thin to overlapping double-layer regime in Fig.~\ref{fig:ChargeVsTime}(b,c), where charging rates are notably slower for the wide and converging cases for the higher resistance entrance, $a_s^*=4$. While the entrance resistance strongly influences charging dynamics, it is not an easily tuned parameter, as it is set by several factors tied to the surrounding electrochemical environment. In practice, the more controllable features lie within the electrode itself---most notably the pore size and pore-size distribution. Because the entrance resistance in our model depends directly on the pore mouth radius, the entrance size provides a practical means to influence the Biot number. This highlights the importance of considering the interface between the electrode and the external electrolyte when evaluating charging behavior; in particular, determining the transport efficiency from the reservoir to the electrode, as it greatly affects the charging rate and which regime is limiting.
\section{Conclusion}
The result outlined in Eq.~\eqref{Eq: diff_eqn_mu_final} provides a differential equation that describes the charging of EDLs in a cylindrical pore with an axially varying pore radius. The reduced-order equation can recover all of the spatiotemporal details, while being 5-6 orders of magnitude faster in computation than the direct numerical simulations. Crucially, we find that a converging conical pore charges faster than both the wider and narrower pores for all Debye lengths. This acceleration occurs due to a boost to both electromigrative and diffusive fluxes as the cross-sectional area decreases. In the limit of thin EDL, {where the analysis and impedance calculations of Keiser et al. hold~\cite{keiser1976abschatzung}}, a favorable electromigration flux direction further accelerates the charging process. Finally, this work highlights the importance of considering entrance resistance, as it can significantly impact electrode charging dynamics.
\par{}
Looking forward, there exist numerous extensions and implications of this work. The most obvious extension is the expand this analysis to non-axisymmetric pores and other shapes, and subsequently integrate these into a pore network model~\cite{henrique2024network} with arbitrary geometries. This will allow one to explore how different shapes interact within complex, interconnected systems---potentially revealing structure-property relationships that do not emerge in the single pore case. For instance, Nguyen et al.~\cite{nguyen2020electrode} recently argued that electrode tortuosity factor may need to be adjusted; our work can allow such effects to be probed while capturing the effects of overlapping double-layers and pore shape changes. Additionally, investigating confinement effects in interconnected pore networks may enable the development of equivalent geometric representations---single pore with an axially varying shape whose charging dynamics mimic those of the full network, allowing for more accurate and simplified models of real-world porous systems.
\par{}
Beyond geometry, incorporating the effects for concentrated electrolytes~\cite{newman2021electrochemical, bazant2009towards}, multi-ion systems~\cite{jarvey2022ion}, diffusivity asymmetry~\cite{henrique2022impact, henrique2025parallel}, Faradaic reactions~\cite{jarvey2022ion, jarvey2023asymmetric}, and ionic liquids~\cite{lee2015dynamics} opens several opportunities for inquiry. Beyond supercapacitors, our work can also advance understanding of electrolyte transport in membranes for separations~\cite{alizadeh2019impact} and biosensing applications~\cite{shah2024universal}. Also, modeling of nanopores has shown potential in capacitive desalination and ion-selective nanopores~\cite{zhang2024addressing, bondarenko2025modelling, wang2024combined}, which shape could further manipulate the dynamics of these systems. Additionally, the response of conical geometries in an AC field has been shown to induce shape-dependent dynamics---particularly relevant in the context of electrokinetic flows~\cite{ratschow2022resonant}-- where our theory could be employed to extend to arbitrary geometries. 
\section{Methods}
\textbf{Numerical solution for the linearized model.} The theoretical framework developed provides the governing equations, boundary, and initial conditions necessary to solve the transient pore charging problem. Using the method of lines, the time-dependent partial differential equations are transformed into a system of ordinary differential equations (ODEs). A finite difference scheme is employed to discretize the spatial domain, and the resulting ODE system is then solved over the specified time domain. \texttt{SciPy}'s built in function \texttt{odeint} was used to solve this ODE system, which uses the Livermore Solver for Ordinary Differential Equations (LSODA). The resulting solution describes the time evolution of the electrochemical throughout the pore. From the electrochemical potential, the electric potential and charge density can be found. Reworking equations \eqref{Eq: mu_wall} and \eqref{Eq: conservationPotential}, the governing expressions for these quantities as functions of radius can be derived. The final expressions for the average charge density, average potential, charge density, and potential are as follows
\begin{subequations}
\label{Eq: rho_phi_from_mu}
    \begin{align}
        \overline{\rho} &= (\hat{\mu} - 2\Phi_w)\frac{2}{\kappa\alpha} \frac{I_1(\kappa\alpha)}{I_0(\kappa\alpha)}  \label{Eq: rho_bar_from_mu}, \\
        \hat{\rho} &= (\hat{\mu} - 2\Phi_w) \frac{I_0(\kappa R)}{I_0(\kappa\alpha)}, \label{Eq: rho_from_rho_bar} \\
        \overline{\Phi} &= \frac{\hat{\mu}}{2} \left( 1 - \frac{2}{\kappa\alpha} \frac{I_1(\kappa\alpha)}{I_0(\kappa\alpha)} \right) + \Phi_w\frac{2}{\kappa\alpha} \frac{I_1(\kappa\alpha)}{I_0(\kappa\alpha)}, \label{Eq: phi_bar} \\
        \hat{\Phi} &= \frac{\hat{\mu}}{2} \left( 1 - \frac{I_0(\kappa R)}{I_0(\kappa\alpha)} \right) +  \Phi_w\frac{I_0(\kappa R)}{I_0(\kappa\alpha)}. \label{Eq: phi}
    \end{align}
\end{subequations}

\textbf{Direct numerical simulation details.} Open-source software OpenFOAM was used to perform direct numerical simulations and solve Eq.~\eqref{Eq: geqn_dim}. A pore of length $10 \ \mu$m and a cylindrical SDL region of length $5 \mu$m, were connected via a transition region with an arc to smooth out the geometrical transition and reduce the numerical errors. The radii of the pores were set to $10$ nm and $20$ nm for the narrow and wide pores, respectively. The converging pore starts from the radius of the wide pore and linearly decreases until it reaches the radius of the narrow nanopore. Conversely, the diverging pore starts with the radius of the narrow pore and linearly increases to the radius of the wider pore. The SDL radii were fixed to be $40$ nm except for the narrow pore cases, in which case the radius was set to $20$ nm. The boundary conditions were kept identical to our prior work of Gupta et al.~\cite{gupta2020PRL}. The vacuum permittivity is $8.85 \times 10^{-12}$ F/m, electrical permittivity of water is $\epsilon=7.1 \times 10^{-10}$ F/m (assuming relative permittivity to be $80.2$), diffusion coefficient $D=1.34 \times 10^{-9}$ m$^2$/s, and reservoir concentration $c_0 = 0.94$ mM were used. The simulations were performed using a 28-core workstation with a computational cost of 4 million seconds for each second elapsed in the simulation.

\begin{acknowledgement}
A.G. thanks the NSF (CBET-2238412) CAREER award for financial support. 
\end{acknowledgement}

\bibliography{acs-achemso.bib}

\end{document}